\documentclass{article}
\usepackage[utf8]{inputenc}
\usepackage{graphicx} 
\usepackage{amsmath}
\usepackage{algorithmicx}
\usepackage[ruled]{algorithm}
\usepackage[noend]{algpseudocode}
\usepackage{xcolor}
\usepackage{amsthm}
\usepackage{url}

\begin{document}


\title{Let's practice what we preach: Planning and interpreting simulation studies with design and analysis of experiments}
\author{Hugh Chipman \\ Acadia University \and  Derek Bingham \\ Simon Fraser University}
\maketitle

\begin{abstract}
Statisticians recommend the Design and Analysis of Experiments (DAE) for evidence-based research but often use tables to present their own simulation studies. Could DAE do better? We outline how DAE methods can be used to plan and analyze simulation studies. Tools for planning include fishbone diagrams, factorial and fractional factorial designs. Analysis is carried out via ANOVA, main-effect and interaction plots and other DAE tools. We also demonstrate how Taguchi Robust Parameter Design can be used to study the robustness of methods to a variety of uncontrollable population parameters.
\end{abstract}



\section{Introduction}

It would be difficult to impossible to find an issue of a major research journal in the statistical and data sciences without a simulation study of some sort. Simulation studies typically generate random realizations of samples from populations, systematically varying population or model parameters across the different realizations. At each combination of parameters, the performance of models, inferential methods or algorithms are evaluated empirically (e.g. significance level, power, predictive ability, etc.). The systematic exploration of parameter combinations enables reporting of results via tables. In other words, a designed experiment is performed on the statistical procedure. 
So, why not use well-established DAE techniques to design simulation studies and synthesize the results?

We illustrate how tools for the design of physical experiments can yield efficient simulation studies, and statistical tools for the analysis of experiments lead to the discovery of easy to interpret results. A systematic and quantitative approach offers the potential of discovering deeper interpretations from simulation studies than would be evident from simply looking at tabulated results. This is especially true for larger studies, where 4, 5 or more factors are varied. In such cases, where the layout of the table could either reveal or hide relationships, it's unrealistic to think that looking at large tables will reveal everything. Tools such as ANOVA are designed for identification and interpretation of important effects.  
We illustrate how design of experiments can save time while giving accurate and intepretable results. For example, rather than run a full factorial design at all combinations of factor levels, fractional factorial designs save computational effort and permit the systematic study of more factors. We also illustrate Taguchi's Robust Parameter Design (TRPD), a design and analysis tool that focuses on identifying and controlling the effects of random variation. In the context of studies of statistical procedures, a TRPD approach helps us find models and methods whose performance is stable across uncontrollable factors such as unknown popluation parameters, sampling schemes and violations of model assumptions. For example, we will use TRPD to guide the choice of a testing procedure and tail direction, so that type I error is stable and as close as possible to the nominal level, across different sample sizes, noise levels and degree of censoring.

Our work is similar in spirit similar to that of Gelman, Pasarica and Dodhia (2002), who argue that tables in statistical papers would be better presented as graphs. Likewise, we argue that statisticians would do well to use their own tools for their own studies.  Our focus is somewhat different than Gelman et al.. They consider a variety of tables, often with different types of measurement in the same table (e.g. both estimator bias and standard error).   Our tables contain a single type of measurement, because of the type of tools we use. DAE tools are usually intended for a ``single response'' case, in which a single measurement is made at different combinations of multiple factors. Fortunately, this type of study is commonplace, making our approach widely applicable.

The paper is outlined as follows. In Section 2, we motivate the problem by introducing a simulation study originally presented in Krishnamoorthy, Mallick and Mathew (2011). Their work examines the effect of 5 different factors on the type I error rate of statistical hypothesis tests. Section 3 presents our general approach to the planning, execution and analysis of a statistical study, based on the five stages of an industrial experimental study. We return in Section 4 to the motivating example, illustrating the five stages in a detailed analysis. The findings of the original study are identified, along with several new results. A limitation of re-analyzing a previous study is that we cannot plan the investigation using DAE tools. With this in mind, Section 5 presents an end-to-end example in which we design, execute and analyze a study of the performance of 2 statistical learning algorithms. We conclude the paper with discussion of some related issues.

\section{Motivating Example}
\label{sec:KrishIntro}

To motivate the application of DAE to the planning and analysis of simulation studies, we consider a previously published study. The point of our illustration is to demonstrate that more can be gained by using the tools of DAE and not to be critical. Krishnamoorthy, Mallick and Mathew (2011, hereafter KMM) examined the performance of 4 hypothesis testing procedures for the mean of a log-normal distribution, using censored data. The performance of left-, right- and two-tailed tests with varying sample sizes were considered. Data were simulated from different populations, varying the proportion of left-censored observations and population standard deviation. Thus a total of 5 factors were varied in the study: {\tt method, tail, n, p0} and {\tt sigma}. 

KMM's results are reported in Table~\ref{tab:Krish_table}. For each entry, 1,000 different realizations of a sample from the lognormal distribution were generated under the null hypothesis. Type I error rates are estimated by the proportion of the 1,000 simulated datasets for which the method rejects $H_0$. KMM make qualitative statements summarizing patterns in the table.

\begin{table}
  \centering
  {\footnotesize 
  \begin{tabular}{rcrrrrrrrrrr}
  & & & \multicolumn{3}{c}{$\sigma = 1$} & \multicolumn{3}{c}{$\sigma = 2$} &  \multicolumn{3}{c}{$\sigma = 3$} \\ 
  n & p0 & method & L & R & T & L & R & T & L & R & T\\ \hline
20 & 0.2 & GV & 5.4 &  4.4 & 5.0 & 6.0 &  5.0 &  5.2 &  4.9 &  4.9 &  5.3\\
 &  & AN & 2.1 & 10.5 & 8.2 & 0.4 & 12.4 &  9.1 &  0.2 & 15.2 & 11.5\\
 &  & SL & 4.6 &  6.9 & 5.9 & 4.3 &  4.0 &  4.0 &  3.5 &  7.7 &  6.2\\
 &  & MS & 5.3 &  4.2 & 4.7 & 5.7 &  3.9 &  4.9 &  6.2 &  3.8 &  5.2\\[1ex]
 & 0.3 & GV & 5.3 &  4.4 & 5.8 & 5.9 &  4.0 &  5.2 &  5.5 &  4.9 &  5.0\\
 &  & AN & 2.1 & 11.0 & 8.3 & 0.4 & 13.7 & 10.2 &  0.1 & 15.1 & 11.7\\
 &  & SL & 4.4 &  5.8 & 5.1 & 3.4 &  6.4 &  5.1 &  4.0 &  6.3 &  5.1\\
 &  & MS & 5.3 &  3.6 & 4.3 & 6.0 &  3.7 &  4.6 &  6.4 &  3.8 &  5.2\\[1ex]
 & 0.5 & GV & 5.9 &  3.6 & 4.4 & 8.0 &  3.3 &  5.3 &  5.5 &  5.4 &  5.5\\
 &  & AN & 2.2 & 10.8 & 8.5 & 0.2 & 15.7 & 12.2 &  0.0 & 17.6 & 13.9\\
 &  & SL & 4.8 &  5.9 & 5.6 & 4.1 &  6.6 &  5.8 &  3.9 &  7.0 &  5.6\\
 &  & MS & 4.9 &  2.9 & 3.6 & 6.0 &  2.8 &  5.1 &  7.8 &  2.7 &  5.3\\[1ex]
 & 0.7 & GV & 6.2 &  1.6 & 4.3 & 7.0 &  2.3 &  5.3 &  5.1 &  4.4 &  4.6\\
 &  & AN & 3.9 &  0.3 & 1.7 & 0.1 & 19.7 & 16.4 &  0.0 & 21.3 & 18.4\\
 &  & SL & 5.2 &  4.0 & 4.3 & 4.3 &  7.0 &  5.6 &  3.3 &  7.3 &  5.4\\
 &  & MS & 4.4 &  3.0 & 3.7 & 7.9 &  1.4 &  4.6 & 10.2 &  1.5 &  6.1\\[1ex]
30 & 0.2 & GV & 5.4 &  4.7 & 5.4 & 5.5 &  4.6 &  4.9 &  6.0 &  4.4 &  5.2\\
 &  & AN & 2.8 &  9.6 & 7.1 & 0.9 & 11.9 &  8.4 &  0.6 & 13.0 &  9.1\\
 &  & SL & 4.6 &  6.0 & 5.5 & 4.3 &  6.5 &  5.6 &  3.7 &  6.6 &  5.2\\
 &  & MS & 5.1 &  4.4 & 4.7 & 5.6 &  4.1 &  4.9 &  5.5 &  4.2 &  4.9\\[1ex]
 & 0.3 & GV & 5.6 &  4.6 & 5.4 & 6.0 &  4.5 &  5.0 &  5.4 &  4.0 &  4.8\\
 &  & AN & 2.5 &  9.2 & 6.9 & 0.9 & 12.6 &  9.2 &  0.3 & 13.8 & 10.4\\
 &  & SL & 5.0 &  6.1 & 5.4 & 4.1 &  6.4 &  5.2 &  3.9 &  6.7 &  5.3\\
 &  & MS & 5.5 &  3.6 & 4.4 & 6.3 &  4.1 &  5.2 &  6.5 &  3.9 &  5.0\\[1ex]
 & 0.5 & GV & 6.0 &  3.9 & 5.2 & 6.2 &  3.3 &  5.0 &  6.1 &  3.3 &  4.3\\
 &  & AN & 2.4 &  9.8 & 7.6 & 0.4 & 14.1 & 10.7 &  0.2 & 14.9 & 11.6\\
 &  & SL & 5.2 &  3.9 & 5.1 & 4.7 &  6.6 &  5.7 &  3.6 &  6.8 &  5.4\\
 &  & MS & 4.8 &  3.3 & 3.9 & 6.4 &  3.4 &  4.7 &  7.3 &  3.3 &  5.2\\[1ex]
 & 0.7 & GV & 6.6 &  2.4 & 4.8 & 7.3 &  4.0 &  5.3 &  7.5 &  2.2 &  4.7\\
 &  & AN & 3.9 &  4.2 & 2.3 & 0.2 & 16.8 & 13.8 &  0.0 & 19.0 & 15.6\\
 &  & SL & 4.5 &  4.0 & 4.2 & 4.0 &  6.2 &  5.3 &  3.6 &  6.9 &  5.2\\
 &  & MS & 4.6 &  2.7 & 3.2 & 7.3 &  1.8 &  4.6 &  8.8 &  1.8 &  5.5\\[1ex]
50 & 0.2 & GV & 5.3 &  5.3 & 4.8 & 5.1 &  5.0 &  4.9 &  5.5 &  4.1 &  4.4\\
 &  & AN & 2.8 &  8.0 & 6.0 & 1.7 & 10.0 &  6.7 &  1.2 & 10.8 &  7.3\\
 &  & SL & 4.3 &  5.9 & 5.5 & 4.3 &  6.0 &  5.4 &  4.3 &  6.3 &  5.6\\
 &  & MS & 5.1 &  4.7 & 4.9 & 5.8 &  4.4 &  5.0 &  5.7 &  4.8 &  5.3\\[1ex]
 & 0.3 & GV & 6.7 &  4.4 & 4.9 & 6.6 &  4.7 &  6.3 &  5.4 &  5.3 &  5.7\\
 &  & AN & 2.6 &  8.3 & 6.0 & 1.4 & 10.1 &  7.1 &  0.8 & 11.5 &  8.1\\
 &  & SL & 4.9 &  5.9 & 5.8 & 4.1 &  5.7 &  5.1 &  4.4 &  5.9 &  5.4\\
 &  & MS & 5.0 &  4.7 & 4.8 & 5.6 &  4.1 &  4.8 &  5.7 &  3.9 &  4.9\\[1ex]
 & 0.5 & GV & 6.1 &  4.1 & 5.2 & 6.6 &  3.2 &  5.1 &  6.2 &  4.4 &  4.9\\
 &  & AN & 2.7 &  8.6 & 6.5 & 0.8 & 12.5 &  9.5 &  0.4 & 12.4 &  9.5\\
 &  & SL & 4.9 &  5.8 & 5.1 & 4.4 &  6.4 &  5.5 &  4.2 &  6.7 &  5.6\\
 &  & MS & 5.0 &  3.9 & 4.3 & 6.2 &  3.6 &  5.2 &  6.2 &  3.7 &  5.1\\[1ex]
 & 0.7 & GV & 7.9 &  1.5 & 4.8 & 8.0 &  4.0 &  5.4 &  7.3 &  2.2 &  5.3\\
 &  & AN & 3.5 &  5.6 & 4.0 & 0.4 & 14.3 & 11.0 &  0.1 & 15.1 & 11.9\\
 &  & SL & 4.8 &  4.5 & 4.7 & 4.2 &  5.8 &  4.9 &  3.7 &  6.6 &  5.2\\
 &  & MS & 4.7 &  4.1 & 4.2 & 6.5 &  2.4 &  4.4 &  7.1 &  2.5 &  4.9\\

  \end{tabular}
  }
  \caption{Estimated type I error rates reported in KMM.  Entries are percentages on a 0 - 100 scale.}
  \label{tab:Krish_table}
\end{table}

The simulation study can be thought of as a full factorial designed experiment with 5 experimental factors ({\tt method} and {\tt tail} of the analysis procedure, {\tt n} of the data collection process and {\tt p0} and {\tt sigma} of the true population model). These factors and their levels are summarized in Table~\ref{tab:Krish_factors}. There are a total of $4 \times 3 \times 3 \times 4 \times 3 = 432$ combinations of {\tt method} $\times$ {\tt tail} $\times$ {\tt n} $\times$ {\tt p0} $\times$ {\tt sigma}. In DAE language, this is a $3^34^2$ full-factorial design with each ``run'' or observation corresponding to 1 of the 432 cells in Table~\ref{tab:Krish_table}.

\begin{table}
  \centering
  \begin{tabular}{c|p{0.25\textwidth}|p{0.5\textwidth}}
     Factor & Description & Levels\\ \hline
     {\tt method} & Hypothesis testing procedure & (AN = \textbf{A}symptotic \textbf{N}ormality of the MLE, SL = \textbf{S}igned \textbf{L}og-likelihood ratio test , GV = \textbf{G}eneralized \textbf{V}ariable method, MS = \textbf{M}odified \textbf{S}igned log-likelihood ratio test)  \\ \hline
     {\tt tail} & Tail of the hypothesis test & L = \textbf{L}eft, R = \textbf{R}ight, T = both \textbf{T}ails \\ \hline
     {\tt n} & Sample size & 20, 30, 50 \\ \hline
     {\tt p0} & proportion of left-censored observations & 0.20, 0.30, 0.50, 0.70\\ \hline
     {\tt sigma} & standard deviation of lognormal population & 1, 2, 3
  \end{tabular}
  \caption{Factors in the type I error study.}
  \label{tab:Krish_factors}
\end{table}

Since the simulation study is a designed experiment, it can be analyzed as such. Table~\ref{tab:KrishANOVA} displays an ANOVA for a 4th order model based on all runs with ``type I error rate'' (recorded as a value between 0 and 100) as the response. The remaining 72 degrees of freedom are treated as residual, although they could be assigned to the 5th order interaction (leaving no estimate of residual error).   Factors {\tt n, p0} and {\tt sigma} are numeric, but treated as categorical in the ANOVA.

\begin{table}
  \centering 
  \footnotesize 
  
\begin{verbatim}
                     Df Sum Sq Mean Sq  F value   Pr(>F)    
method                3  555.1   185.0  629.713  < 2e-16 ***
tail                  2  332.3   166.1  565.361  < 2e-16 ***
n                     2   11.4     5.7   19.353 1.88e-07 ***
p0                    3    2.7     0.9    3.117 0.031319 *  
sigma                 2   99.2    49.6  168.785  < 2e-16 ***

method:tail           6 2258.0   376.3 1280.705  < 2e-16 ***
method:n              6   50.4     8.4   28.571  < 2e-16 ***
method:p0             9   35.4     3.9   13.374 2.64e-12 ***
method:sigma          6  137.7    23.0   78.107  < 2e-16 ***
tail:n                4   11.6     2.9    9.867 2.02e-06 ***
tail:p0               6   21.7     3.6   12.290 1.79e-09 ***
tail:sigma            4   90.1    22.5   76.693  < 2e-16 ***
n:p0                  6    1.0     0.2    0.545 0.771926    
n:sigma               4   11.9     3.0   10.139 1.45e-06 ***
p0:sigma              6   80.7    13.5   45.777  < 2e-16 ***

method:tail:n        12   48.3     4.0   13.693 3.39e-14 ***
method:tail:p0       18   60.1     3.3   11.370 1.54e-14 ***
method:tail:sigma    12  298.1    24.8   84.525  < 2e-16 ***
method:n:p0          18    5.0     0.3    0.936 0.539622    
method:n:sigma       12   13.4     1.1    3.813 0.000169 ***
method:p0:sigma      18  103.3     5.7   19.528  < 2e-16 ***
tail:n:p0            12    0.9     0.1    0.268 0.992458    
tail:n:sigma          8    5.1     0.6    2.170 0.039827 *  
tail:p0:sigma        12   37.3     3.1   10.591 1.08e-11 ***
n:p0:sigma           12   10.8     0.9    3.050 0.001627 ** 

method:tail:n:p0     36    6.8     0.2    0.640 0.928695    
method:tail:n:sigma  24   18.4     0.8    2.606 0.000950 ***
method:tail:p0:sigma 36  129.9     3.6   12.281  < 2e-16 ***
method:n:p0:sigma    36   22.2     0.6    2.098 0.003836 ** 
tail:n:p0:sigma      24    6.1     0.3    0.866 0.644092    

Residuals            72   21.2     0.3                      
---
Signif. codes:  0 ‘***’ 0.001 ‘**’ 0.01 ‘*’ 0.05 ‘.’ 0.1 ‘ ’ 1
\end{verbatim}
\caption{ANOVA table for 432-run full-factorial experiment.}
  \label{tab:KrishANOVA}
\end{table}

Even a cursory inspection of the ANOVA table indicates many significant effects, including high-order interactions. For example, the 2-way interaction {\tt method:tail} is very large, indicating that type I error rates vary with both factors, and that the way in which the error rates vary across tail type varies by method. Such interactions are difficult to spot by inspecting tabulated results, and often relies on how the factors are arranged in rows and columns as in Table \ref{tab:Krish_table}.  In Section~\ref{sec:KrishFull} we demonstrate how ANOVA gives insight into the study.

\pagebreak

\section{The five stages of a simulation study}
The example in the previous section highlights that more can be learned from a simulation study than merely inspecting a table of the quantities of interest (QoIs) from each simulation. Instead, we propose using the tools.

Broadly speaking, an experiment consists of five stages: (i) problem definition, (ii) planning, (iii) execution, (iv) analysis and (v) conclusions (e.g. MacKay \& Oldford 2000). Keeping in mind that downstream stages can impact upstream decisions, the stages are very much connected. For instance, the choice of QoI in the problem definition impacts the analysis of the simulation experiment, but the choice of analysis method impacts the simulation design in the earlier planning stage. 

While not identical to physical experiments, the stages can be adapted to simulation studies. We describe this for very generic simulation studies, but the ideas can be adapted to almost any setting. The goal (i) is frequently to assess the performance of methods that combines the simulation results into QoIs that allow for comparison (e.g., power, integrated mean square error, confidence interval coverage, etc.).  Comparing methods is not the only possible objective.  For example, the goal may be comparing the performance of estimators using an appropriate QoI. If the comparison to be made is between using the sample mean and the sample median for the estimator of the population mean, the QoI may be mean square error of each estimator over the simulation runs.   We will return to this running example at several points in this section.

\begin{figure}
  \centering
  \includegraphics[width=0.7\textwidth]{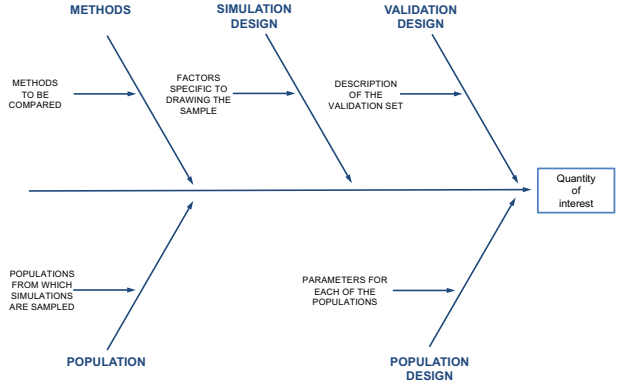}
  \caption{Cause-and-effect diagram.}
  \label{fig:fishbone}
\end{figure}

Planning (ii) a simulation study is trickier. A key early task is to establish the list of factors to include in the experiment.  A large number of potential factors should be identified, and then reduced.  A reduced set of factors will lead to an experimental design that can be run in a reasonable amount of time. A useful tool for organizing the factors that might impact a QoI is a cause-and-effect diagram. In physical experiments, the different factors impacting the performance of a process are usually grouped as branches on the cause-and-effect diagram.  These branches correspond to the major categories: methods, machines, people, materials and measurement. In a simulation study for comparing methods, the factors that one might consider adjusting can instead be categorized under methods, population, population design, simulation design and validation design as depicted in Figure \ref{fig:fishbone}.  In a standard cause-and-effect diagram the population design settings would usually be given by sub-structures within the population design. We have found it easier to visualize by separating the two.

Looking at Figure \ref{fig:fishbone}, it is straightforward to fill in some of the branches. Under methods, for example, the list of statistical methods to be compared is likely clear from the context. This might be, for example, a proposed method and other competing methods.  Similarly, \emph{Population} is a factor when there is more than one population model from which simulations are to be drawn (e.g., data arising from a normal population or a t-distribution). The population design is a little more complicated. Under this branch, the population parameters (e.g., mean structure, effect size, variance and correlation amongst the predictors and responses) are potential factors. Together, the population and population design branches determine the distributions from which the data are simulated . 

Facing any experimenter is the issue of sampling from the population. The simulation design consist of factors that are external to the population such as the size of the sample and covariate settings. Of course, there are grey areas. For example, fixed covariates are likely to placed under the simulation design. If random covariates are being used, the covariate distributions may be specified under the population design.  The validation design typically would entertain similar factors as generating the simulation data, but may consider other population parameters if robustness is being investigated, or different covariate settings if extrapolation is being studied. 

The cause-and-effect diagram is a first step at identifying and organizing the factors, but typically lists more factors than can be accommodated in most physical experiments. The factors thought to be of most interest are then selected from the factors listed on the cause-and-effect diagram, while the remaining factors are fixed at nominal levels in subsequent stages. We recommend doing the same.

At this point, one is faced with the selection of the settings of each factor, or factor levels, to include in the simulation study. For some factors, this task is straightforward. If \emph{model} is a factor, then the factor levels are simply the list of models included in the study. On the other hand, for a factor such as the magnitude of a regression coefficient or an error variance, the choice in levels need be realistic in real-world settings. Furthermore, the levels settings for each factor should be sufficiently distant so that potentially meaningful differences in the QoI can be observed.

\begin{figure}
  \centering
  \includegraphics[width=0.7\textwidth]{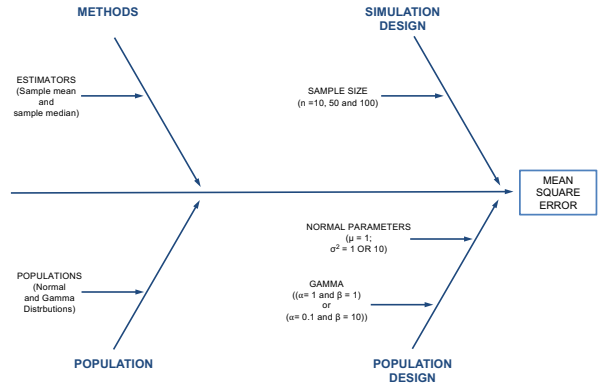}
  \caption{Cause-and-effect diagram for comparing the sample mean and sample median as estimators of the population mean.}
  \label{fig:fishbone-mean-meadian}
\end{figure}

Looking back to the example of estimating the population mean with the sample mean and median, the number of factors is likely not too large and the cause-and-effect diagram is fairly simple (e.g., Figure \ref{fig:fishbone-mean-meadian}). Estimators are the only factors under the Methods branch.  The sampling populations are chosen to be normal and gamma distributions (Population branch).  Under the Population Design branch we have a fixed mean, and variances of 1 and 10.  Also explored (under the Simulation Design branch) is the sample size (n = 10, 50 and 100).  This example is not a predictive model, so there are no covariates, or any other factors under the Validation Design branch.

The choice of factor level combinations to include in the simulation corresponds to choosing the experimental design.  For example, in the mean vs. median example, a full factorial design could be formed with all $2 \times 2 \times 2 \times 3$ possible combinations of the levels of the 4 factors.  
If there are $q$ factors, each with only two levels, there are $2^q$ possible level setting combinations (i.e., runs) for a full factorial design. As the number of factors and the number of levels
grows, the number of runs per replicate of the simulation design increases quickly. If there are ten factors, each with two levels, there are 1024 possible runs for each replicate of the simulation study. 
A study this large can be too time-consuming or expensive to execute.
When full factorial designs cannot be run, experimenters typically turn to fractional factorial designs (e.g., Box, Hunter \& Hunter 2005). We propose the use of fractional factorial experiments for simulation studies as well. In the illustrations considered in this paper, we consider only two-level fractional factorial designs, however there is a rich literature on fractional factorial designs with more than two levels and also different numbers of levels for each factor (Wu \& Hamada 2009; Mukerjee \& Wu, 2006).

With the choice of level settings and factor level combinations (i.e., the simulation experimental design) in hand, it is tempting to move directly to stage (iii) and execute the entire simulation study. Before doing so, there are still some details to address. First, one must decide how many replicates of the simulation design to conduct. Indeed, it is likely that the size of the simulation design and the number of replicates are chosen together. Second, it is frequently worth running a small pilot study to see if the simulations can be executed as planned. The pilot study also allows evaluation of the choice of each factor levels settings and develops an estimate of the required computational resources to complete the study. For example, if interest lies is comparing the mean and the median as estimators of the population mean and the population variances considered were 1 and 2 and the sample sizes were also small, it is likely that the differences would be too small to reach concrete conclusions.  A small pilot study (in which a wide range of sample sizes were considered while holding other factors fixed) would likely detect this and lead to different choices of factor level settings.

Following the pilot study, the simulation design is run. This is a conceptually simple step insofar as the desired simulations are performed. A single instance of the simulation may include simulating data for estimating a model, simulating a validation set, model fitting, making predictions or conducting hypothesis tests, and computing the QoI. If the tasks to be performed are relatively fast, then the simulation study may be done in a sequence. On the other hand, a simulation may involve inversion of large matrices, challenging optimization, many steps of a MCMC, or combinations of these. In such cases, it will be necessary to parallelize the simulation runs.

The analysis stage (iv) is the study of the relationship between the QoI and the factors under study. As with the analysis of designed experiments, we recommend ANOVA models and associated tools be the primary techniques used in this stage. The usual ``statistical toolbox'' for analysis, including model diagnostics, hypothesis testing, transformation, prediction intervals, etc, will be useful. Indeed, the purpose of a careful simulation design is to make the analysis efficient, interpretable and also relatively easy to do.

Later in the analysis, we also illustrate TRPD as a modelling and optimization tool for settings in which some factors are usually under the control of researchers, such as the choice of inferential procedure or model, and other factors are uncontrollable, such as an unknown population variance. In TRPD, we seek combinations of the ``control'' factors that give results that are robust across the levels of the uncontrollable (or ``noise'') factors.  In the case of comparing estimators of the population mean, the estimators would be control factors, and the population distribution would typically be a noise factor. A TRPD analysis would focus on the ``best" estimator, but also identify the estimator that was more robust to changes in population distribution.
In terms of the categories for our cause-and-effect diagrams, it is often the case that factors on the Method branch will be control factors, and factors under the other branches will be noise factors.  For an introductory overview of TRPD, see Nair et al. (1992) and references therein. 

The conclusions (v) should flow naturally from the analysis. A convenient feature of ANOVA models is the ability to identify important factors and, in the case of factors with more than two levels, determine which factor levels are significantly different. For significant main effects and interaction effects, we recommend using plots to observe the magnitude and nature of the effects.

The aim of this paper is to propose that the tools outlined above be applied to the design of simulations studies. In the following sections, we illustrate how the five experimental stages can aid in understanding the performance of methods in two examples.


\section{Return to motivating example}
\label{sec:KrishFull}

We revisit the KMM study, considering the five stages of experimentation outlined in Section 3.  One complete cycle of the stages is demonstrated in Section~\ref{sec:Krish5stages}, including a TRPD analysis.  Additional analysis of higher-order interactions in Section~\ref{sec:KrishHigherOrder} gives further insights.  Our conclusions are compared to the original findings in the KMM paper in Section~\ref{sec:KrishCompare}.  We end by demonstrating in Section~\ref{sec:KrishCheapo} that similar conclusions could have been reached by a smaller experiment.

\subsection{The five stages in the KMM example \label{sec:Krish5stages}}

\subsubsection*{(i) Problem definition:} The aim of the KMM study was to investigate the merits of four hypothesis testing procedures for the lognormal mean, in the presence of left-censoring, using simulation. As such, the problem definition is identified, and the chosen QoI for the simulation is the type I error rate.  Ideally, the type I error is close to the nominal 5\%. As in KMM, we will assume that it is preferable to be at or slightly below the nominal level. So, a 4\% type I error is preferable to 6\% because it controls the type I error below the nominal level of 5\%. 

\subsubsection*{(ii) Planning:} 
\begin{figure}
  \centering
  \includegraphics[width=0.7\textwidth]{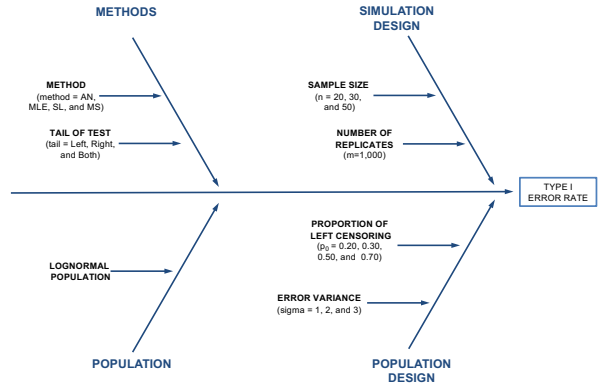}
  \caption{Cause-and-effect diagram for the KMM simulation study.}
  \label{fig:KMM}
\end{figure}

Since we did not participate in the planning of the experiment, a full description this stage is impossible. In particular, a full list of potential variables to include in the experiment is not available. The cause-and-effect diagram for the full $3^34^2$ full factorial design in the simulation study is presented in Figure \ref{fig:KMM}. 
We include a few factors that were held fixed, but KMM would likely been considered as potential factors, such as the distributional family (lognormal) and population mean (set at 0). 
For each factor level combination, 1,000 simulations were drawn from the lognormal distribution under the null hypothesis with a nominal type I error rate of 5\%. 

\subsubsection*{(iii) Execution}

We would normally recommend a small pilot study to investigate whether the level settings are likely to result in meaningful differences in the QoI.  
The pilot study allows the experimenters to estimate the run time of computations for the full suite of simulations.  In the KMM study, computation time was probably not a major hurdle.

If a pilot study had been carried out, it is quite possible that the {\tt AN} method would have been identified as having much larger variation in type I error than any other method. Reading across any {\tt AN} row of Table~\ref{tab:Krish_table} reveals much more variation in type I error for this method than any other method.  A pilot study, looking a small subset of the full table would be likely to identify the poor performance of {\tt AN}.

\subsubsection*{(iv) Analysis}
In light of the terrible performance of the {\tt AN} method, we remove the 108 runs with {\tt method = AN}, and analyze the remaining 324 runs as a $3^44^1$ full factorial design.  In the event that a pilot study did not identify {\tt AN}'s terrible performance, it would show up in a preliminary analysis.  For example, the main effects plot and 2-way interaction plots (not shown) indicate the poor type I error performance of the {\tt AN} method, including a mean type I error rate of 7.6\%.

It is common to analyse simulation studies by visual inspection of the tabulated QoI (e.g. Table~\ref{tab:Krish_table}).  Detection of all but the largest effects can be time-consuming and depends heavily on how the table was constructed (i.e., which factors are placed on columns and rows, and also which are inner and outer column factors).  Instead, we propose using ANOVA, or related methods, for the analysis of the simulations.

Table~\ref{tab:KrishANOVA2} shows a summary of an ANOVA for the 4th order model using the 324 runs. In comparison to the ANOVA of the 432 runs (Table~\ref{tab:KrishANOVA}), fewer interactions are now significant, and the estimate of residual variance (MSE, the Mean Square for residuals) is considerably smaller than for the original 4th order ANOVA (original MSE = 0.29, new MSE = 0.19).  

\begin{table}
  \centering
  
  \footnotesize 
  
\begin{verbatim}
                     Df Sum Sq Mean Sq F value   Pr(>F)    
method                2  11.34    5.67  29.152 5.16e-09 ***
tail                  2  52.96   26.48 136.192  < 2e-16 ***
n                     2   0.59    0.30   1.521 0.228797    
p0                    3   4.07    1.36   6.973 0.000548 ***
sigma                 2   8.52    4.26  21.913 1.73e-07 ***

method:tail           4 215.13   53.78 276.611  < 2e-16 ***
method:n              4   0.46    0.11   0.588 0.672858    
method:p0             6   1.56    0.26   1.337 0.259441    
method:sigma          4   7.46    1.86   9.590 8.81e-06 ***
tail:n                4   0.54    0.13   0.694 0.599625    
tail:p0               6  43.31    7.22  37.123  < 2e-16 ***
tail:sigma            4   1.18    0.30   1.518 0.211780    
n:p0                  6   0.84    0.14   0.719 0.635852    
n:sigma               4   2.41    0.60   3.102 0.023800 *  
p0:sigma              6   9.12    1.52   7.818 6.76e-06 ***

method:tail:n         8   7.58    0.95   4.875 0.000194 ***
method:tail:p0       12  16.33    1.36   6.998 4.00e-07 ***
method:tail:sigma     8  35.23    4.40  22.651 7.55e-14 ***
method:n:p0          12   4.57    0.38   1.957 0.050457 .  
method:n:sigma        8   2.47    0.31   1.587 0.153641    
method:p0:sigma      12   3.87    0.32   1.660 0.106537    
tail:n:p0            12   0.50    0.04   0.214 0.997052    
tail:n:sigma          8   2.08    0.26   1.335 0.249705    
tail:p0:sigma        12   0.99    0.08   0.423 0.946606    
n:p0:sigma           12   1.50    0.12   0.642 0.795650    

method:tail:n:p0     24   4.54    0.19   0.974 0.513875    
method:tail:n:sigma  16   7.83    0.49   2.517 0.006996 ** 
method:tail:p0:sigma 24  21.98    0.92   4.710 2.51e-06 ***
method:n:p0:sigma    24   6.93    0.29   1.485 0.120822    
tail:n:p0:sigma      24   2.52    0.10   0.539 0.948027    

Residuals            48   9.33    0.19                     
---
Signif. codes:  0 ‘***’ 0.001 ‘**’ 0.01 ‘*’ 0.05 ‘.’ 0.1 ‘ ’ 1
\end{verbatim}

\caption{ANOVA table for 324-run full-factorial experiment excluding {\tt method = AN}.}
  \label{tab:KrishANOVA2}
\end{table}

\subsubsection*{Main effects and two-factor interactions}

\begin{figure}
  \centering
  \includegraphics[width=\textwidth]{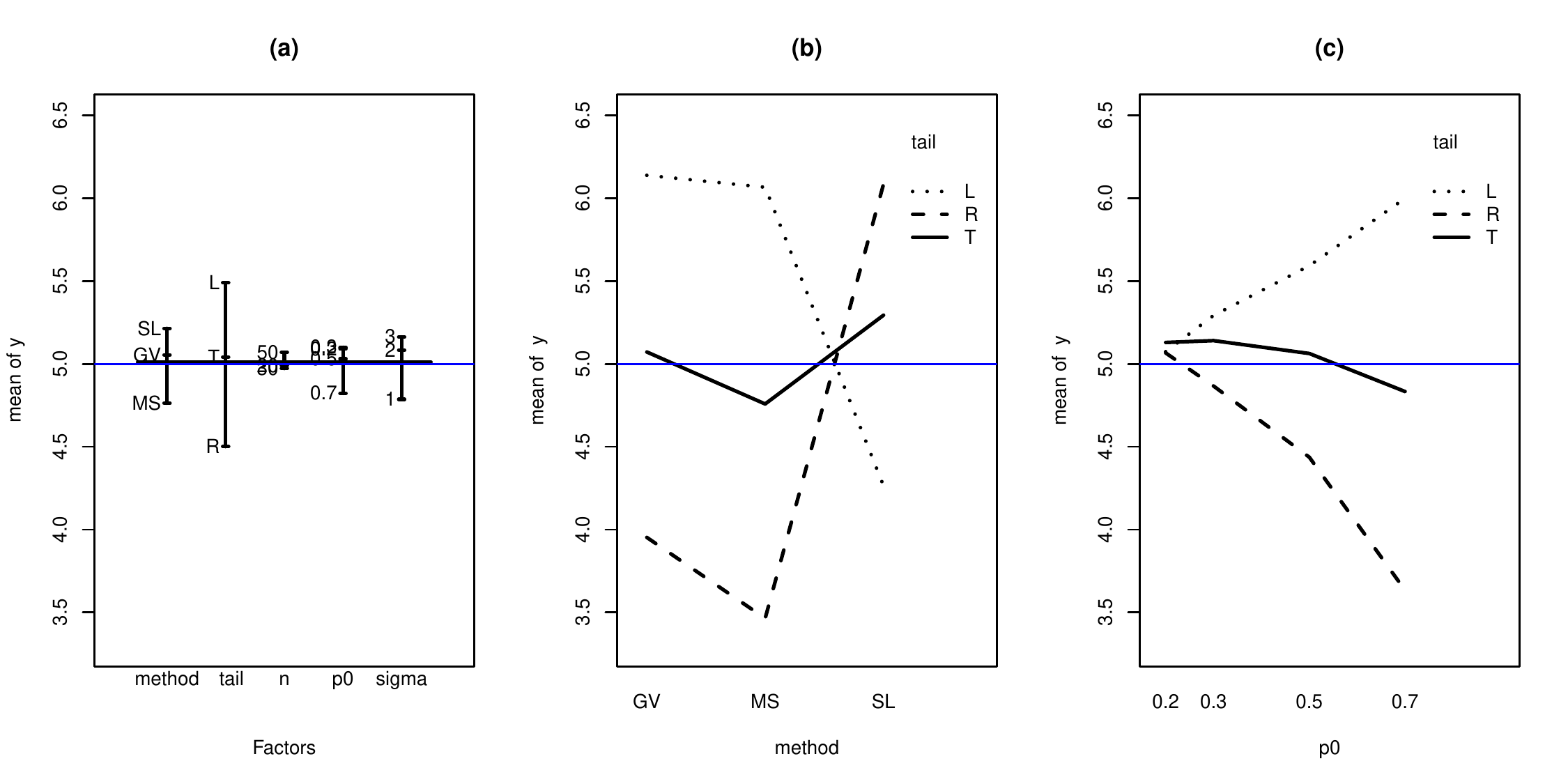}
  \caption{Estimated main effects (a) and select 2-factor interactions (b), (c). The vertical axis is the response, ``Type I error rate'', which should have a nominal level of 5\%, indicated by the horizontal blue line. }
  \label{fig:Krish_me2fi}
\end{figure}

Main effects are displayed in Figure~\ref{fig:Krish_me2fi}(a). Of note are:
\begin{itemize}
  \item The largest main effect is {\tt tail}.
  \item Main effect magnitudes are small, and the main effect of {\tt n} is negligible.
  \item Response means are close to the nominal level of $\alpha=5\%$. 
  \item For the numeric factors {\tt n, p0, sigma}, effects are monotone or almost so. The most visible example is increasing mean response as {\tt sigma} increases from 1 to 3.
\end{itemize} 

It might be tempting to conclude from the main effect plot that since main effects are small, the type I error is always close to the desired 5\%.  However when interpreting effect plots, it's important to bear in mind that they show one effect at a time. The full prediction from the model is {\em additive} in main effects and interactions. Thus we proceed to look at 2-way interactions.

The two largest two-factor interactions, {\tt method:tail} and {\tt tail:p0} are plotted in Figure~\ref{fig:Krish_me2fi}(b) and (c). While other interaction effects are significant, they are much smaller. Large interaction effects are seen as non-parallel lines. 

The largest two-factor interaction is {\tt method:tail} (Figure~\ref{fig:Krish_me2fi}(b)). All 3 methods exhibit considerable range in type I error rate over the 3 different tails. For example, when {\tt method=MS}, the mean type I error rate is just above 6\% for a left-tailed test, while it is well below 4\% for a right-tailed test. Thus, a change in {\tt tail} from left to right produces an (interaction) effect for the {\tt MS} method of about 2.5\%. This effect will be additive with the effects involving other terms, and could push predictions of type I error rate much higher or lower. 

Figure~\ref{fig:Krish_me2fi}(c) shows the {\tt tail:p0} interaction effects. The type I error rate for left- and right-tailed tests varies with censoring proportion {\tt p0}. The downward-sloping dashed line indicates that right-tailed tests become increasingly conservative as {\tt p0} increases. The nearly flat solid line indicates that for two-tailed tests, type I error rate is stable and close to 5\%. Ideally, the choice of method would be one that results in a type I error rate close to 5\% across changes of factors that are not under the control of the experimenter. This desire for robustness leads us to consider the analysis of the KMM study in the context of TRPD.

\subsubsection*{Taguchi Robust Parameter Design}

The interpretation above of the {\tt tail:p0} interaction suggests that the two-tailed tests are robust to changes in {\tt p0}. The idea of TRPD is to identify experimental factors that can be controlled, and choose levels of those factors so that the response has a value equal to a target, with small variation. 

Here is a quick primer on TRPD in this setting:
\begin{itemize}
  \item The factors {\tt method} and {\tt tail} are control factors, which can be chosen by an analyst.
  \item The factors {\tt n}, {\tt p0} and {\tt sigma} are noise factors, which are largely beyond the control of the analyst. One might argue that {\tt n} can be controlled, but for illustrative purposes, we will assume that samples are expensive, and so changing {\tt n} is unlikely to be controllable.
  \item The target value for the response in this example is the nominal 5\% level.
  \item The idea of TRPD is to choose a level of the control factors so that responses are close to the target and are robust (insensitive) to the uncontrollable (noise) factors. Large interactions between control and noise factors provide an opportunity to achieve such robustness.
\end{itemize}
The previously identified {\tt tail:p0} interaction in (Figure~\ref{fig:Krish_me2fi}(c)) can be interpreted in the context of TRPD. The {\tt tail} of the hypothesis test is the control factor since the user could choose that. The user cannot control censoring level {\tt p0}, making it a noise factor. We want to choose a {\tt tail} so that no matter what {\tt p0} is, our Type I error rate is close to the desired 5\%. Choosing {\tt tail = T} gives a more robust response (a line closer to horizontal).

\begin{figure}
  \centering
  \includegraphics[width=0.8\textwidth]{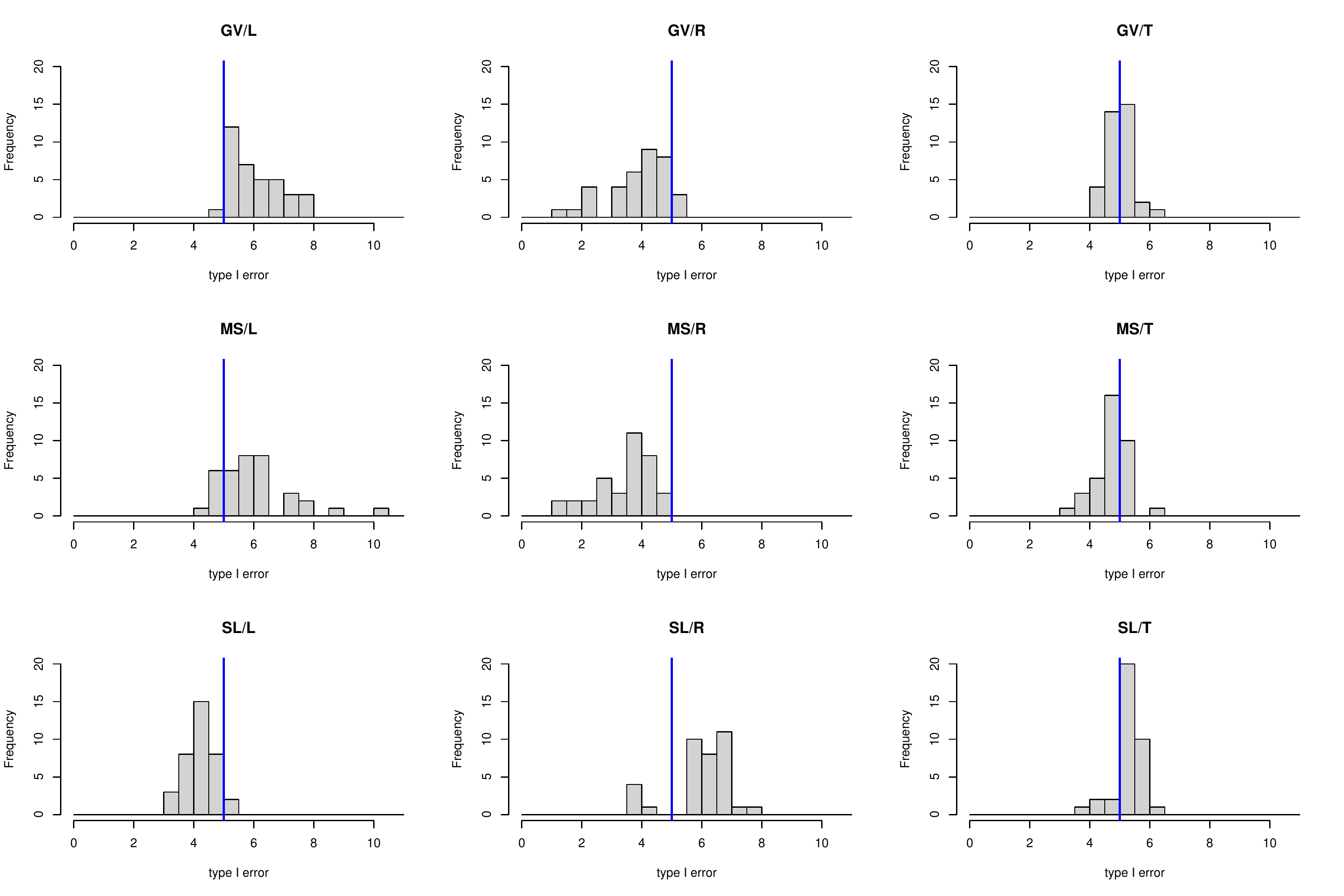}
  \caption{Histograms of response within each combination of {\tt method:tail}. Columns of the grid correspond to {\tt tail} and rows correspond to {\tt method}, as indicated by titles above each histogram.}
  \label{fig:Krish_TaguchiHist}
\end{figure}

Another way to examine robustness is to ``lump together'' the noise effects and look at the variation in response at each combination of the 2 control factors {\tt method} and {\tt tail}. Figure~\ref{fig:Krish_TaguchiHist} does this, showing histograms of the response for each of the 9 combinations of {\tt method} and {\tt tail}. We see that:
\begin{itemize}
  \item 2-tailed tests give type I error closer to the desired 5\%, no matter what method is used. The histograms in the right column are centered around 5\% and have small variation.
  \item If a left-tailed test is desired, the {\tt SL} method is preferred over the other methods. It is slightly conservative with levels just below 5\% and less variation in level than the other methods. The other methods are too liberal and have some levels well above 5\%.
  \item If a right-tailed test is desired, {\tt GV} and {\tt MS} are conservative (sometimes quite so), while {\tt SL} is liberal. All exhibit considerable variation in type I error, but the conservative tests may be preferred.
\end{itemize}

Notice that the findings relate to both the centre and spread of the distribution of response values. Choice of control factor levels as a way of controlling both a target (i.e., location) and variation is a central element of TRPD.

\subsubsection*{(v) Conclusions}
The TRPD analysis has identified some important conclusions, in addition to the previous conclusions that the {\tt AN} method is terrible, and some 2-way interactions are larger than the fairly small main effects.  We postpone further discussion of conclusions until Section~\ref{sec:KrishCompare}.

\subsubsection*{The end? Not in this case}
Having identified and examined main effects and two-way interactions and considered robustness, an analysis would often be complete at this point. Higher-order effects are often small.  We have illustrated the five stages.  In this example, however, it's worthwhile to press on further. Higher-order interactions appear large (ANOVA in Table~\ref{tab:KrishANOVA2}). Comparing the two-way interaction plots (Figure~\ref{fig:Krish_me2fi}) and the range of type I errors (extremes visible in Figure~\ref{fig:Krish_TaguchiHist}), we see the range exceeds the mean levels in the interaction plot.  This also suggests that higher order interactions may explain additional variation in type I error.

\subsection{Higher-order analysis and TRPD again \label{sec:KrishHigherOrder}}

The presence of significant and large 3rd and 4th order effects (ANOVA in Table~\ref{tab:KrishANOVA2}) indicates that further analysis beyond main effects and 2-way interactions is necessary. As in the last section, we analyze the subset of 324 runs with {\tt method $\neq$ AN}. Visualizations need some modification to simultaneously display the effect of 3 or more factors on the response.

To visualize 3-way interactions, we {\em combine} the two ``control'' factors {\tt method} and {\tt tail} into a single factor with 9 {\tt method/tail} levels ({\tt GV/L, GV/R, GV/T, MS/L, ..., SL/T}) and plot the new combined control factor against the 3rd factor and the response. The layout is similar to 2-way interaction plots in Figure~\ref{fig:Krish_me2fi}, but with 9 lines for the 9 control factor combinations. Line type and color represent {\tt method} and {\tt tail} respectively. Such a modified plot isn't always required for TRPD, but the prevalence of higher-order interactions in this example makes it necessary.

\begin{figure}
  \centering
  \includegraphics[width=0.8\textwidth]{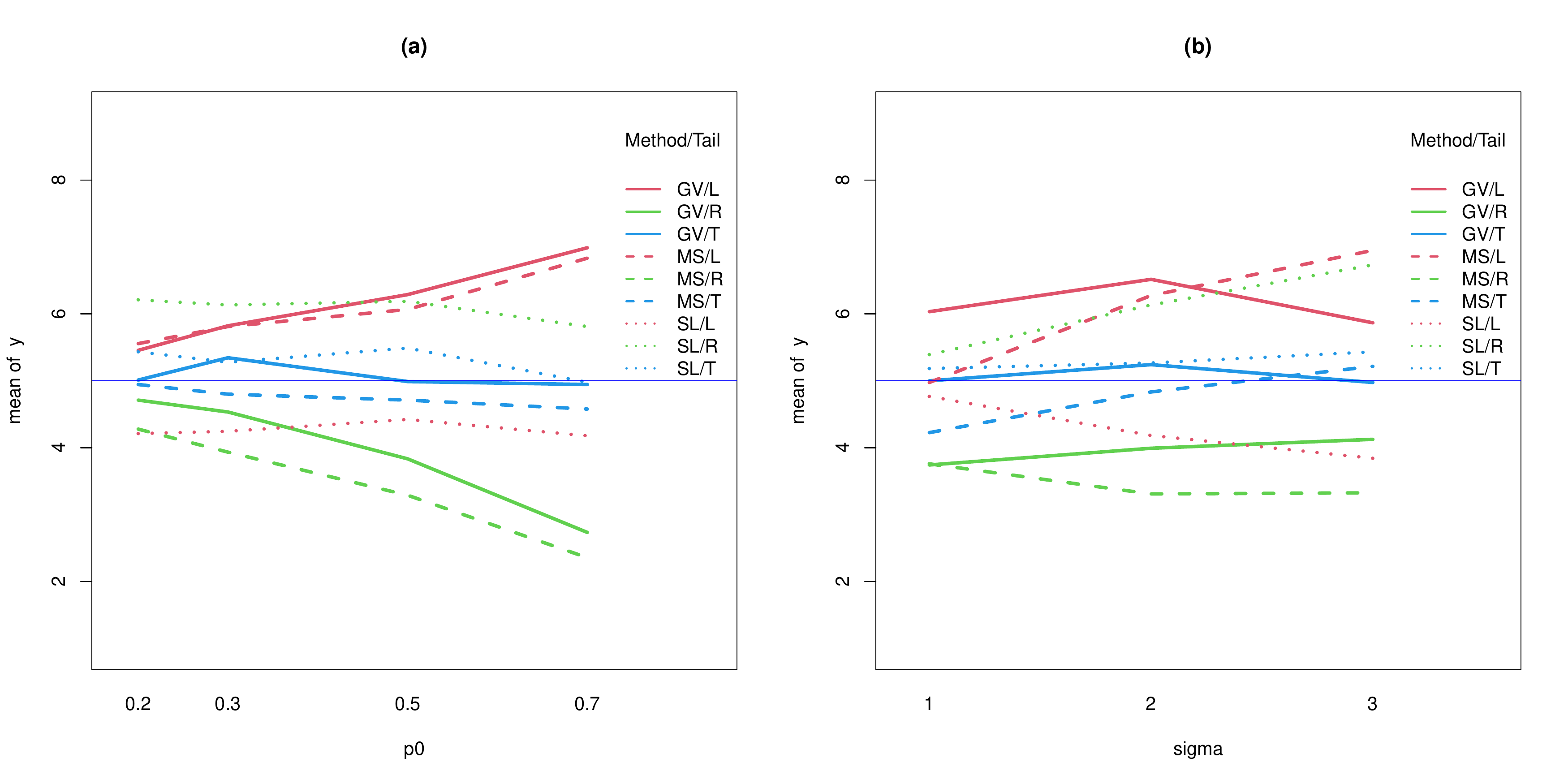}
  \caption{Three-factor interaction plots for TRPD, showing interaction effects {\tt method:tail:p0} (a) and {\tt method:tail:sigma} (b). Line type represents {\tt method} and line colour represents {\tt tail}. }
  \label{fig:Krish_Taguchi3fi}
\end{figure}

The previous ANOVA table (Table~\ref{tab:KrishANOVA2}) identifies the two largest three-factor interactions as {\tt method:tail:p0} and {\tt method:tail:sigma}.  These interactions are visualized in Figure~\ref{fig:Krish_Taguchi3fi}. 
The fact that control by noise interactions are large suggests that TRPD can be used to identify sources of uncontrollable variation, giving more insight into the methods. As mentioned earlier, the idea of TRPD is that we want to choose a level of the control factor so that responses are robust (insensitive) to the uncontrollable noise factors. That is, we want to choose a {\tt method} and {\tt tail} so that no matter what {\tt n}, {\tt p0} and {\tt sigma} are, our Type I error rate is close to the desired 5\%.

Some interpretations of these results (not previously seen in the histograms):
\begin{itemize}
  \item From the {\tt method:tail:p0} interaction plot (Figure~\ref{fig:Krish_Taguchi3fi}(a)), both the {\tt GV} and the {\tt MS} methods have poor type I error for large {\tt p0} (i.e. large amounts of censoring). The solid and dashed lines have an upward slope for left-tail (red) and downward slope for right-tail tests (green).
  \item From the {\tt method:tail:p0} interaction plot (Figure~\ref{fig:Krish_Taguchi3fi}(a)), the {\tt SL} method does not change its level as censoring changes (dotted lines are horizontal). However, the left-tailed SL test (red dotted) is conservative and the right-tailed SL test (green dotted) is liberal.
  \item From the {\tt method:tail:sigma} interaction plot (Figure~\ref{fig:Krish_Taguchi3fi}(b)), the {\tt MS} and {\tt SL} methods have better type I errors for {\tt sigma=1} than {\tt sigma = 2} or {\tt 3}. The {\tt GV} method seems less sensitive to {\tt sigma}.  That is, the 3 solid lines are closer to horizontal.
\end{itemize}

One final issue to bear in mind when interpreting the interaction plots is that they only show a portion of the full ANOVA contributions of all 5 factors to the response. For example, Figure~\ref{fig:Krish_Taguchi3fi}(a) shows a {\tt method:tail:p0} interaction. The predicted response may shift up or down, depending on effects involving {\tt sigma} and {\tt n}. The analysis we carry out focuses on the plots one at a time, effectively averaging over other factors.

In the KMM study, some fourth-order effects are also large. These are studied in Supplementary Materials.


\subsection{Comparison to the findings of KMM }
\label{sec:KrishCompare}

KMM summarize their findings of this study over two paragraphs of text. Below, we quote their findings and compare them to our conclusions ({\em in italic}). We then identify some additional findings not given in KMM.

\begin{enumerate}
  \item ``The test based on the asymptotic normality of the MLE (denoted by AN in Table 1) seems to be the worst among all tests....Thus, the test AN should be avoided in applications.'' {\em This was overwhelmingly clear and we dropped runs corresponding to {\tt method=AN} in our analysis. A pilot study might have identified and eliminated {\tt AN} before carrying out the full experiment.  }
  \item ``Performance depends on {\tt sigma}, {\tt p0} and {\tt tail}.'' {\em We observe this also, although we additionally note that the dependence is complex with higher-order interactions that would be difficult to see by visual inspection of the results. Some effects containing {\tt n} are statistically significant, but are relatively small contributions.}
  \item ``No test emerges as the clear winner in all the scenarios.'' {\em We come to the same conclusion (e.g., see Figure~\ref{fig:Krish_TaguchiHist}).}
  \item ``For left-tailed testing, the SLRT appears to exhibit the most satisfactory performance, even though it is conservative in some cases. Both the GV approach and the MSLRT appear to be quite liberal in most cases, especially when {\tt sigma} is large.'' {\em Figure~\ref{fig:Krish_TaguchiHist} also confirms all these conclusions.}
  \item ``For right-tailed testing, it is the GV approach that provides the most satisfactory solution.'' {\em {\tt GV} is good for right tail tests, but so is {\tt MS} (Figure~\ref{fig:Krish_TaguchiHist})}.
  \item ``For two-tailed testing, all the three procedures provide satisfactory performance, even though the SLRT is liberal and the MSLRT is conservative in a few cases. The GV approach appears to be quite satisfactory for two-sided testing, for all the cases considered in the simulation.'' {\em Our findings are similar.}
  \item ``As the sample size gets large, there is very little difference between the GV approach and the MSLRT. In several instances, the MSLRT is not as satisfactory as the SLRT.'' {\em Table 3 indicates that only one significant effect involves {\tt n}, namely {\tt method:tail:n}. KMM's finding is based partly on a seperate run not included in the study, with {\tt n = 100}. Over the range {\tt n = 20, 30, 50}, the effect of sample size is relatively small. As previously noted, we would expect that the relatively small effect of {\tt n} might have been detected in a pilot study, and potentially suggested including n=100 in places of one of the other settings (likely n=30).}
\end{enumerate}

Did we discover anything KMM did not? Perhaps:
\begin{itemize}
  \item Type I error rates vary in a complex way over the 5 factors. This is evidenced by large and significant interaction effects, even at 3rd and 4th order.
  \item After excluding {\tt method = AN}, left tail tests are, on average a bit liberal. Right tail tests are a bit conservative, and two-tailed tests are on average at the correct level. 
  \item Sample size {\tt n} has a relatively small effect, compared to the other factors. This may be in part due to the range of values chosen for {\tt n}. Indeed the authors report some additional results for {\tt n = 100}, not contained in the main study.
  \item KMM identify only {\tt GV} as the preferred {\tt method} for right-tail tests. We find that {\tt method = MS} is just as good (Figure~\ref{fig:Krish_TaguchiHist}).
  \item By systematically analyzing the response, we were able to identify the important factors. In this case, it turned out that there were significant and important high-order interactions, many involving {\tt method} and {\tt tail}. While this complicated the interpretation somewhat, the end result was an analysis in which we are confident we have found the most important effects.
  \item The small effects involving sample size {\tt n} could be due to a rather small range of factor levels (20 to 50). One might hope that with sufficiently large samples, type I error rates would approach the nominal level, as asymptotics begin to play a role. 
  \item The presence of large interactions between the 4 factors {\tt method}, {\tt tail}, {\tt sigma} and {\tt p0} (see Supplementary Materials) suggest that some combinations of {\tt method} and {\tt tail} may be more stable across uncontrollable factors. As {\tt sigma} and {\tt p0} increase, one-sided tests deviate more from the nominal level (``cone shape'' of Y vs. {\tt p0} in Figure~\ref{fig:Krish_Taguchi4fi} (Supplementary Materials), with more deviations for large {\tt sigma}).
\end{itemize}

\subsection{A ``cheapo'' analysis of the type I error study \label{sec:KrishCheapo}}

Can you learn the same thing with less? That is, could a smaller number of runs have been used to reach similar conclusions? Two ways to reduce the number of runs are i) using fewer levels for the factors and ii) using a fractional factorial design rather than considering all combinations of the runs - both reminiscent of issues in DAE. Here we consider reducing the number of levels (i).  See Section~\ref{sec:super1} for an illustration of analysis of a fractional factorial.

For numeric factors with more than 2 levels, we use only the extreme levels ({\tt n = 20} or {\tt 50}, {\tt p0 = 0.2} or {\tt 0.7} and {\tt sigma = 1} or {\tt 3}). We also restrict attention to the 3 ``better'' methods, excluding {\tt method = AN}. This reduces the original table with 432 entries (324 after discarding {\tt method = AN}) to $3\times 3 \times 2 \times 2 \times 2 = 72$ runs. 

An ANOVA for this reduced data is shown in Table~\ref{tab:KrishANOVA3}. There are fewer significant effects than in the analysis of the 324 runs in Table~\ref{tab:KrishANOVA2}. However, the five largest effects (by SS) are the same as in Table~\ref{tab:KrishANOVA2}: {\tt tail}, {\tt method:tail}, {\tt tail:p0}, {\tt method:tail:sigma} and {\tt method:tail:p0:sigma}. All are significant in the reduced analysis.

Analysis of the reduced 72 runs uncovers many of the same patterns. The main conclusions are the same, namely that {\tt n} is not an important factor, but that (some) interactions between the other 4 factors are important and that there are high-order interactions. Figures~\ref{fig:Krish_me2fi} - \ref{fig:Krish_Taguchi3fi} are re-done for the 72 runs, and presented in the Supplementary Materials. The findings are generally the same. 

One could use a fractional factorial design to further reduce the number of simulations. However for this study, 5 factors is a smallish number and computations are quite fast, leaving little reason to do so. Additionally, the choice of fractional factorial designs for mixed-level data is a somewhat advanced topic. 

\begin{table}
  \centering
  
  \footnotesize 
  
\begin{verbatim}
                     Df Sum Sq Mean Sq F value   Pr(>F)    
method                2   3.01   1.507  15.178 0.013556 *  
tail                  2  11.94   5.968  60.094 0.001037 ** 
n                     1   0.01   0.011   0.113 0.753350    
p0                    1   3.25   3.251  32.740 0.004617 ** 
sigma                 1   3.69   3.690  37.159 0.003663 ** 

method:tail           4  51.55  12.888 129.786 0.000174 ***
method:n              2   0.56   0.282   2.836 0.171010    
method:p0             2   0.43   0.215   2.165 0.230581    
method:sigma          2   1.67   0.834   8.397 0.037002 *  
tail:n                2   0.18   0.090   0.910 0.472202    
tail:p0               2  18.26   9.128  91.917 0.000453 ***
tail:sigma            2   0.22   0.110   1.109 0.413803    
n:p0                  1   0.19   0.190   1.915 0.238649    
n:sigma               1   1.05   1.051  10.586 0.031269 *  
p0:sigma              1   2.46   2.457  24.740 0.007630 ** 

method:tail:n         4   6.40   1.600  16.108 0.009851 ** 
method:tail:p0        4   3.32   0.829   8.350 0.031872 *  
method:tail:sigma     4  19.22   4.804  48.375 0.001214 ** 
method:n:p0           2   0.87   0.437   4.403 0.097571 .  
method:n:sigma        2   0.44   0.222   2.232 0.223323    
method:p0:sigma       2   0.17   0.087   0.878 0.482816    
tail:n:p0             2   0.10   0.052   0.522 0.629043    
tail:n:sigma          2   0.41   0.203   2.043 0.244667    
tail:p0:sigma         2   0.10   0.051   0.513 0.633251    
n:p0:sigma            1   0.59   0.587   5.909 0.071914 .  

method:tail:n:p0      4   3.20   0.799   8.045 0.033968 *  
method:tail:n:sigma   4   3.10   0.776   7.813 0.035707 *  
method:tail:p0:sigma  4   9.85   2.463  24.799 0.004391 ** 
method:n:p0:sigma     2   0.68   0.341   3.429 0.135694    
tail:n:p0:sigma       2   0.14   0.069   0.698 0.549551    

Residuals             4   0.40   0.099                     
---
Signif. codes:  0 ‘***’ 0.001 ‘**’ 0.01 ‘*’ 0.05 ‘.’ 0.1 ‘ ’ 1
\end{verbatim}
\caption{ANOVA table for reduced 72-run full-factorial experiment, excluding {\tt method = AN}.}
  \label{tab:KrishANOVA3}
\end{table}


\section{Example: Planning a Statistical Learning simulation study ``from scratch''}

In this section, the five stages of experimentation are illustrated using a study that investigates the predictive accuracy of two statistical learning models with a continuous response. The purpose of the illustration is to highlight how the five stages can be considered in practice. A fractional factorial design is introduced as a way to reduce the number of runs.

\subsection{The five stages, with a fractional factorial experiment \label{sec:super1}}

\subsubsection*{(i) Problem Definition}

Our study examines predictive accuracy of possible models for supervised learning. A variety of response variables could be used to measure predictive accuracy. We will focus on $R^2$, the percent variation explained, evaluated on large ($n=10,000$) test sets. 

\subsubsection*{(ii) Planning}

\begin{figure}
  \centering
  \includegraphics[width=0.7\textwidth]{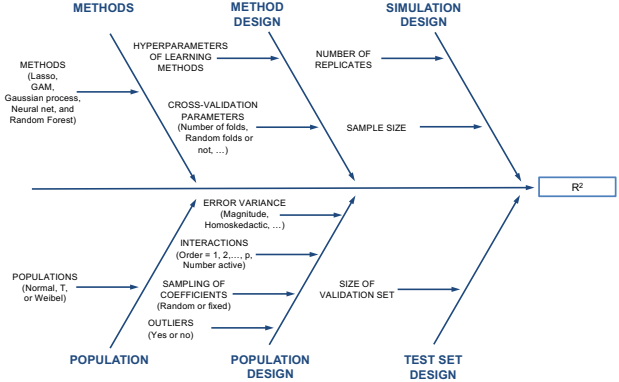}
  \caption{Cause-and-effect diagram for the statistical learning simulation study.}
  \label{fig:SL}
\end{figure}

To begin, a cause-and-effect diagram (Figure \ref{fig:SL}) is constructed to organize the list of potential factors. While one can imagine many potential factors to include, an experiment based solely on those in the Figure \ref{fig:SL} can get quite large. Some choices have to be made to keep the run size manageable. To keep things simple, for example, two statistical learning procedures (lasso and random forests) are compared and only a normal parent population is used. 
We eventually chose seven factors, each with two levels, and excluded or held fixed all other factors (e.g. we used only normal errors and did not investigate outliers).   The factors are summarized in Table~\ref{tab:SL_factors}.

\begin{table}
  \centering
  \begin{tabular}{c|p{0.45\textwidth}|p{0.3\textwidth}}
     Factor & Description & Levels\\ \hline
     {\tt model} & Statistical learning model & lasso, random forest \\ \hline
     {\tt n} & Training sample size & 250, 1000 \\ \hline
     {\tt q} & Number of measured variables & 20, 50\\ \hline
     {\tt ENE} & Expected number of active effects & 10, 20 \\ \hline
     {\tt beta.mu} & Coefficient values of active effects & 1, 3 \\ \hline
     {\tt sigma} & Error standard deviation & 0.5, 2 \\ \hline
     {\tt x.cor} & Correlation parameter for predictors & 0, 0.8 \\ \hline
  \end{tabular}
  \caption{Factors in the statistical learning study. See text for details.}
  \label{tab:SL_factors}
\end{table}

We now discuss the choice of levels for each of the factors. In all simulations, training and test data are generated from a linear model with normal errors,
\begin{equation}
  y = X'\beta + \varepsilon, \qquad \varepsilon \sim IID\ \ N(0,\sigma^2), \label{eq:linear}
\end{equation}
where $X$ is a $p-$vector of predictor variables formed from $q$ measured variables ($q < p$). (``Measured variables'' is used to identify distinct variables that can be thought of as each being ``measured''; predictor variables are constructed from the ``measured variables''). The predictors in the linear model will include linear terms equal to individual measured variables and product terms formed by multiplying together 2 measured variables. Thus there are a total of $p = q + \binom{q}{2} = q(q+1)/2$ predictor variables. Some elements of the coefficient vector $\beta$ are exactly 0, corresponding to inactive predictors.

The two statistical learning models are :
\begin{enumerate}
  \item Least squares regression using the lasso and potential predictors consisting of all possible linear effects and product effects (total of $q(q+1)/2$ predictors).
  \item Random forest regression using $q$ measured variables as predictors. 
\end{enumerate}
Since the form of the lasso model matches the data-generating mechanism, we expect it to outperform random forests.

Both procedures have tuning parameter(s) that must be chosen. In the case of lasso regression, there is a single penalty parameter $\lambda$ which will be chosen by running 10-fold cross-validation within the training set. In the case of random forests, three potential tuning parameters are {\tt mtry}, the proportion of predictors that are randomly sampled when growing each branch of a tree, {\tt ntree}, the number of trees in the random forest ensemble, and {\tt maxnodes}, the maximum number of nodes that a tree can contain. Although there are 3 tuning parameters, pilot runs indicated that {\tt ntree = 500} and no limit on {\tt maxnodes} gave the best results. The remaining random forest parameter, {\tt mtry}, will be chosen from a grid of proportion values (0.05, 0.10, 0.20, 0.30, 0.40, 0.50, 0.60, 0.70, 1), using the out-of-bag data. Thus, both lasso regression and random forests are using sample-reuse techniques to choose the optimal tuning parameter.

The sample size of the training set is {\tt n}. The nonzero elements of $\beta$ will all be equal to {\tt beta.mu}. The error standard deviation {\tt sigma} corresponds to $\sigma$ in (\ref{eq:linear}). The $q$ ``measured variables'' are simulated as multivariate normals with mean vector 0 and covariance matrix $\Sigma$. The diagonal elements of $\Sigma$ are always 1, giving measured variables that are standard normal variates. The factor {\tt x.cor} controls the correlation between the measured variables, with the correlation between variables $i$ and $j$ given by
$$
\Sigma_{i,j} = \Sigma_{j,i} = \exp(\log(\mbox{\tt x.cor})|i-j|).
$$
This decaying correlation gives a (maximal) correlation of {\tt x.cor} between measured variables with adjacent indices (e.g. $j=i+1$), and smaller correlation for other pairs of measured variables. This formulation of the measured variables results in ``product terms'' having mean 0 and variance 1, so that scaling of the predictors is not needed.

Some elements of the regression coefficient vector $\beta$ are set to zero. Rather than having a fixed pattern, this is done at random. An ``effect heredity'' prior distribution (Chipman 1996) specifies a prior probability that a main effect will be active as $\pi$, assuming that activity of main effect are independent of each other. Conditional on whether ``parent'' main effects $A$ and $B$ are active, the probability that the $AB$ product term is active is given as
\begin{equation} \label{GEN_PRIOR_2fi}
P(AB\mbox{ active } | \mbox{ activity of } A,B)=\left\{\begin{array}{lll}
        0 & \mbox{if both $A, B$ inactive}\\
        c_1\pi & \mbox{if exactly one of $A, B$ active}\\
        c_2\pi & \mbox{if both $A, B$ active}.
                \end{array}\right. .
\end{equation}
For given values of $q, \pi, c_1, c_2$ the expected number of active main effects, active 2-way interactions with 2 active parents, and active 2-way interactions with 1 active parent can be calculated. In our experiment, $\pi, c_1$ and $c_2$ are chosen so that:
\begin{itemize}
  \item A total of {\tt ENE} effects are expected to be active.
  \item {\tt ENE}/2 main effects are expected to be active.
  \item {\tt ENE}/4 2-way interactions with 2 active parents are expected to be active.
  \item {\tt ENE}/4 2-way interactions with 1 active parents are expected to be active.
\end{itemize}

The factor {\tt ENE} is set at a value chosen as part of the experimental design. For given values of {\tt ENE} and {\tt q}, and using the 50/25/25 split of active effects specified above, the parameters $\pi, c_1, c_2$ take unique values and can be easily calculated. Derivations are sketched in the Supplementary Materials. In each simulation run, the actual number of active effects will be a draw from this distribution. The division of the active effects among main effects, single-parent 2-factor interactions and two-parent 2-factor interactions will also be random.

Instead of running a full factorial $2^7$ design (with 128 runs), we introduce fractional factorial designs as a way of saving computational effort. In this example, we will choose a design with 32 runs. Fractionating a 2-level full factorial design is a well-studied problem and is a standard topic in introductory DAE texts (e.g. Box, Hunter \& Hunter 2005). A 2-level fractional factorial is typically constructed by selecting rows based on ``generators'', which are identities that apply to the columns of the design matrix. To simplify notation we label the 7 factors in our experiment as {\tt n} = A, {\tt q} = B, {\tt ENE} = C, {\tt beta.mu} = D, {\tt sigma} = E, {\tt x.cor} = F and {\tt model} = G, and assume each 2-level factor is coded with levels $-1$ and $+1$. The generators we will use are ABCE = +1 and BCDF = +1. Of the 128 runs of the full-factorial (without replication), 32 runs satisfy both these conditions, making this a quarter fraction of the full factorial design. For these runs, the generators will imply aliases between effects. That is, an effect estimate will actually correspond to the sum of several effect estimates. For example, the AB interaction will be aliased with the CE interaction (implied by the first generator). The AB interaction will also be aliased with other higher-order terms. The generators imply the following aliasing patterns in this case:
\begin{enumerate}
  \item Main effects A - E are each aliased with 2 different three-way interactions, and higher order terms.
  \item Main effect G is aliased with terms of order 5 and higher
  \item Two-way interactions that do not involve G are aliased in pairs, e.g. AB = CE.
  \item Two-way interactions involving G are aliased with four-way interactions and higher-order interactions.
\end{enumerate}
The presence of aliasing between pairs of two-way interactions implies that this is a resolution IV design.

This particular fractional factorial was chosen because the factor {\tt model} (G) is of special interest as the only control factor in TRPD. Similar to the analysis of the KMM study in Section~\ref{sec:KrishFull}, control by noise interactions give insight into the robustness of the models with respect to uncontrollable factors. The generators chosen enable estimation of the main effect for G and two-way interaction effects involving G, provided we assume that fourth-order and higher 
interactions are negligible. This does come at a cost, namely the inability to disentangle two-way interactions not involving G.

\subsubsection*{(iii) Execution}

Pilot studies were used to prototype the code and check for reasonable factor levels. Random Forests required more compute time, so during prototyping we temporarily set {\tt ntree = 50}.  Table \ref{tab:SL_factors} contains the levels settings for the final experiment. However, the original choices for the level of the training sample size were $n=1,000$ and $n=10,000$. It turned out that both sample sizes were large enough that non-meaningful differences in the response variable was observed. Instead, we changed the levels shown in Table \ref{tab:SL_factors}.  

The 32-run quarter fraction design was executed.  Every run was a completely independent draw from the population model. Runs were parallelized, using 2 cores of an 8-core laptop and taking under 30 minutes to execute.

\subsubsection*{(iv) Analysis}

The response analyzed (QoI) was a logistic transform of the test $R^2$. If {\tt y.test} is the vector of observed (i.e. simulated) $y$ values in (\ref{eq:linear}) and {\tt yhat.test} is the corresponding vector of model predictions for an independent test set, then we define $R^2=\mbox{cor}({\tt y.test},{\tt yhat.test})^2$ and analyze as our response $$\log \left( \frac{R^2}{1-R^2} \right).$$ Preliminary analysis of an untransformed $R^2$ as the response gave unsatisfactory residual diagnostics and predictions of $R^2$ that exceeded 100\% in some cases. The ANOVA model used here is based on least-squares / normal theory, rather than being a GLM. 

An advantage of using 2-level designs is that the simpler half-normal plot (Daniel 1959) can replace an ANOVA table as the primary analysis technique. In balanced designs like the ones we have discussed, all effects are estimated by contrasts with equal variance. Effect estimates for terms that are inactive will all be normally distributed with mean 0. The half-normal plot is a quantile-quantile plot of absolute effect estimates against corresponding quantiles of a half-normal distribution (i.e. the absolute value of a standard normal). Effects that are small will fall on a straight line, while large effects will be larger than this line (to the right of the plots in our figures). 

\begin{figure}
  \centering
  \includegraphics[width=0.55\textwidth]{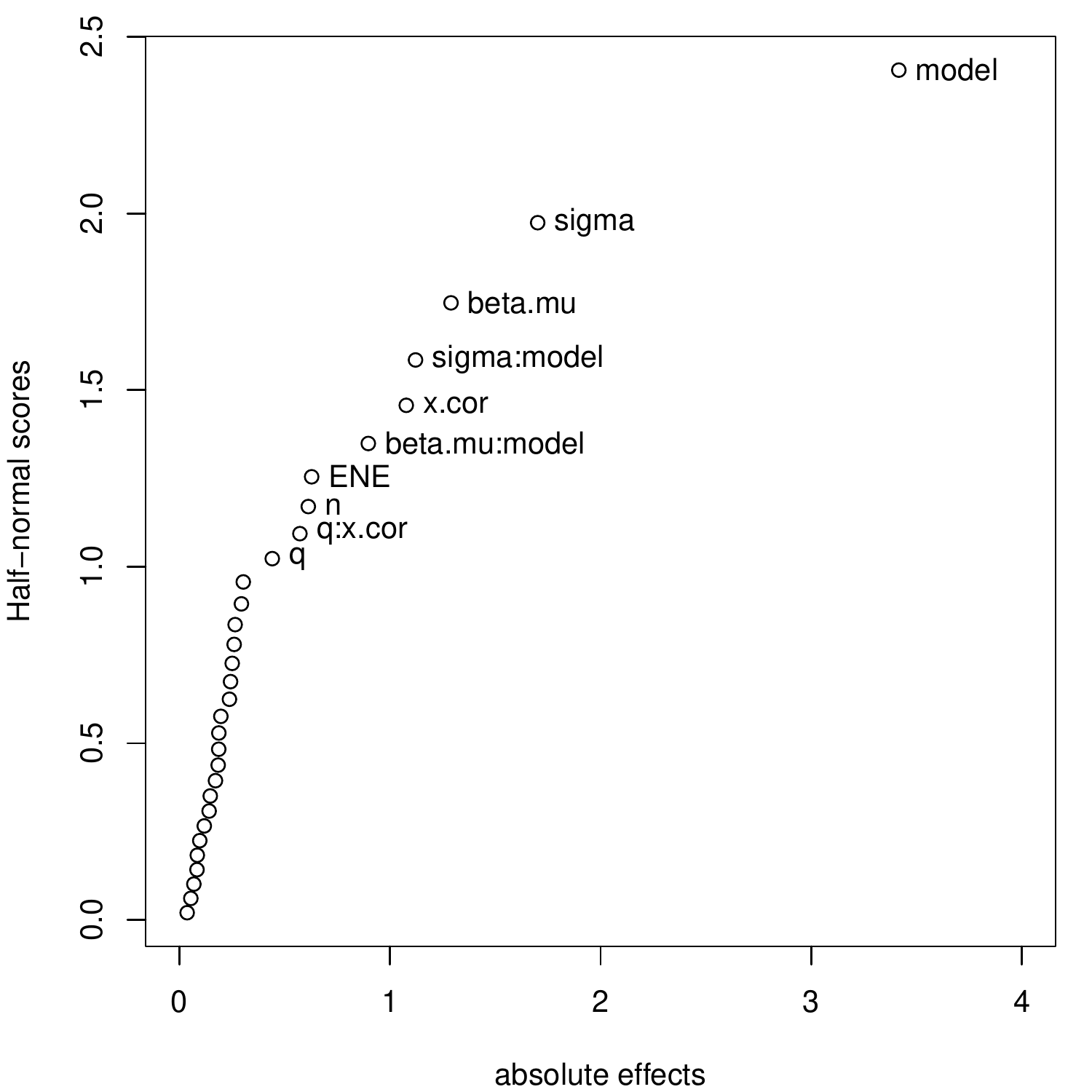}
  \caption{Half-normal plot of estimated effects, statistical learning experiment.}
  \label{fig:superhalfnormal}
\end{figure}

The half-normal plot in Figure~\ref{fig:superhalfnormal} suggests that between 6 and 10 effects are important. The large effects are labelled while the remaining smaller, unlabelled, effects appear to form a linear trend. The 6 largest effects are {\tt model, sigma, beta.mu, sigma:model, x.cor} and {\tt beta.mu:model}. The 4 large main effects are aliased with 3-way interactions, while the 2 interactions involving {\tt model} are aliased with 4-way interactions. 

\begin{figure}
  \centering
  \includegraphics[width=\textwidth]{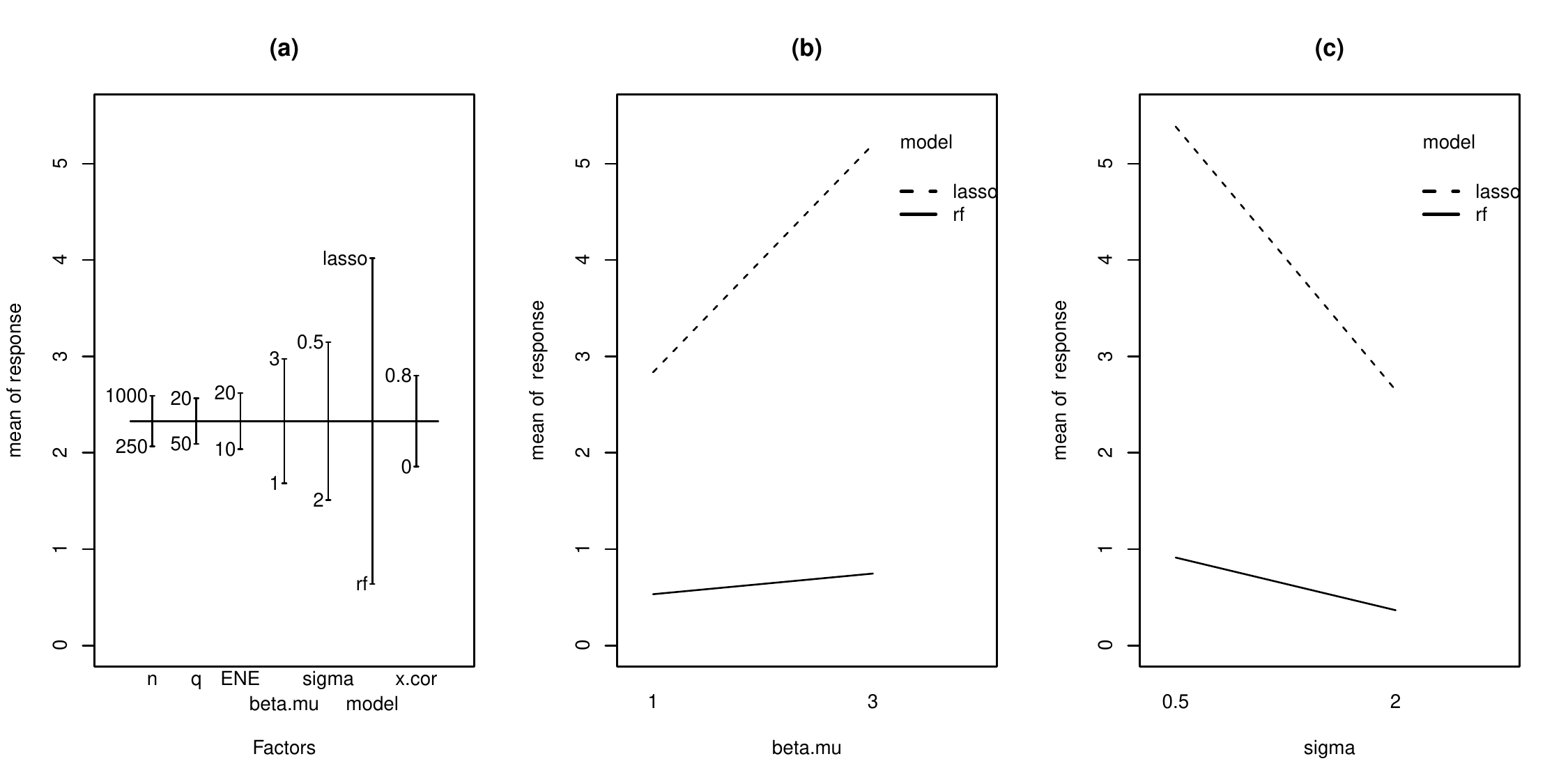}
  \caption{Main-effect plot (a) and select interaction plots (b,c) for supervised learning experiment.}
  \label{fig:superME2FI}
\end{figure}

The dominant effect is {\tt model}, with other large effects including {\tt beta.mu}, {\tt sigma}, {\tt x.cor}, and interactions {\tt beta.mu:model} and {\tt sigma:model}. These effects are represented visually in Figure~\ref{fig:superME2FI}(a) and (b) - (c).

Although the response is a transformation of $R^2$, it is still a ``larger the better'' measure. With this in mind, the largest main effects and their interpretations are:

\begin{itemize}
  \item {\tt model}, with lasso performing much better than random forests.
  \item {\tt sigma}, with better predictive accuracy when the training set has less noise ({\tt sigma = 0.5}).
  \item {\tt beta.mu}, with better predictive accuracy when the coefficients are larger ({\tt beta.mu = 3}).
  \item {\tt x.cor}, with better predictive accuracy when the measured variables are correlated ({\tt x.cor = 0.8}). This is likely because the test set has the same correlation patterns among the measured variables, and the performance measure is predictive accuracy, not selection of the correct subset of predictors.
\end{itemize}

In the interaction plots, the largest effects involve {\tt model} and one other term. Keeping in mind that a large interaction will appear as a large change in {\em slope}, rather than mean level, the interaction effects and their interpretations are:

\begin{itemize}
  \item {\tt beta.mu:model}, with lasso models seeing a larger gain in predictive accuracy than random forests, when signal levels are high ({\tt beta.mu = 3}). That is, the slope of the lasso line is much steeper than random forests.
  \item {\tt sigma:model}, with lasso models seeing a larger loss in predictive accuracy than, when noise levels are high ({\tt sigma = 0.5}). That is, the slope of the lasso line is negative and larger than the slope of the random forests line.
\end{itemize}

\subsubsection*{(v) Conclusions}

In this experiment, the lasso is the clear winner. This should not be surprising since the form of the population model matches the lasso regression. Considering TRPD, interaction plots (Figure~\ref{fig:superME2FI}(b), (c)) show less variation in the performance of random forests than in the lasso, across levels of the noise factors {\tt beta.mu} and {\tt sigma}. But even though the lasso model has more variable performance, it is always superior to random forests. The better performance (higher $R^2$) of lasso dominates the increased sensitivity to noise factors, and so for this study TRPD turns out to be a secondary consideration.

\subsection{Other remarks}

Other experimental designs and their analysis were considered but not reported here. These include a $2^7$ full factorial with 128 runs and 2 or more replicates of the same design (with 256 or more runs). Supplementary materials demonstrate that for this application, the conclusions would be effectively the same.

Fractionated designs will imply some effects are aliased. In this case, the aliasing did not impact the conclusions. In cases where aliasing was a problem, additional runs could be later added to reduce aliasing due to fractionation. 

A different fractional factorial could have been chosen, with fewer aliases between 2-way interaction terms. Doing so would treat effects more symmetrically, rather than prioritizing effects involving the only control factor, {\tt method} (labelled G). For example, the generators ABCF = +1 and ABDEG = +1 give a resolution 4 design, but with just 6 out of 21 two-way interactions aliased with a two-way interaction (compared to 15 of 21 in the design we chose).

If TRPD considerations are important (e.g., we want to study robustness to uncontrollable factors), the simplest fractional factorial designs to analyze are {\em crossed arrays}, in which designs are constructed separately in the control factors and in the noise factors, and then the designs are ``crossed''. That is, for every run in the control-factor design, all runs in the noise factor design are carried out. One or both of the crossed designs could be fractionated. In this example we have 6 noise factors, and our design is a crossed array, choosing a quarter fraction of a design in the noise factors, and running this design once for G=1 (lasso) and once for G = -1 (random forests).

\section{Conclusion}

This is not rocket science.  The tools we have used would usually be covered in an undergraduate course in DAE, with perhaps the exception of TRPD.  We hope that readers will decide that it's obvious to design and analyze statistical studies using statistical tools, that the process of doing so is straightforward, and that there are real benefits to doing so.  

A systematic approach like this formalizes some of the steps that would normally be carried out, such as identifying relevant factors to consider and appropriate levels (both in the planning stage).  But other parts are (sadly) novel: Even if a full-factorial experiment is run (planning stage), an ANOVA combined with graphical exploration of effects should make it easier to discover all the important effects (analysis stage).  In circumstances where there are many factors that are important to consider in the study, fractional factorial designs (planning, again) offer the possibility of exploring them all, rather than artificially eliminating some factors in order to design a study that can be run.  The TRPD framework (planning and analysis) formalizes the common objective that when choosing among the models being compared, it's desirable to pick a method that is stable (i.e. robust) across the scenarios we cannot control.

There are several considerations and open problems not discussed in the paper that we mention briefly.\\[-20pt]

\subsubsection*{Randomization restrictions to reduce noise, blocking and random effects}

Our treatment (and both examples) implicitly assume that the random errors are independent and identically distributed across runs.  For example, in the statistical learning study, every run in our design drew a separate, independent realization of data from the population model.  Thinking further about that example, a natural temptation would be to ``pair'' together runs that are the same except for the learning method used (lasso vs. random forests).  So the 32 runs used in our example could be grouped into 16 pairs, with 16 different datasets (rather than 32).  Both methods would be applied to the 16 datasets, yielding 32 runs.  One might argue that it would be a ``fairer'' comparison of the 2 models to have them analyze exactly the same data. 

Thinking about this structure, this would be equivalent to adding 16 blocks to the design.  Such a design could be analyzed via block terms or a random effects model.  Essentially we would be analyzing the performance difference between the 2 models as a function of the other 6 factors in the study.  

Randomization restrictions will complicate analysis.  When might such restrictions be worthwhile?  It seems intuitive that it would be worthwhile when the restrictions reduce the error variance, enabling better estimation of effects with the same computational resources.  In our specific statistical learning example, the variation in performance could be mostly explained by the experimental factors, and so pairing or other randomization restrictions were not worth pursuing.  If there was more unexplained variation due to random ``dataset to dataset'' variation there would be a stronger case to employ randomization restrictions to reduce the impact of this variation.

Randomization restrictions can be made more general than the  example above.  If you are doing a TRPD and have a crossed array design, a row of the noise factor design gives you a realization of simulated data.  You can use the same realization of simulated data across all rows of the control factor design.  Another way to generalize could include nested restrictions.  For example, in our statistical learning example, for every realization of data, we simulate a new noise realization $\varepsilon$ in (\ref{eq:linear}).  Instead, we might have grouped together runs so that only {\tt sigma} changed within each group.  Then generate 1 realization of $\varepsilon$ instead of 3, and just multiply $\varepsilon$ by the value of {\tt sigma} to {\em scale} the noise.\\[-20pt]

\subsubsection*{Calibration / fair competition / reproducibility} 
The studies typically involve the comparison of multiple methods.  It's important that the methods be on an equal footing.  This would include choice of tuning parameters, prior distributions (for a Bayesian model), and so on.  For example in the statistical learning study, we took pains to identify the pertinent tuning parameters ($\lambda$ for the lasso, and {\tt mtry} for random forests) and use reasonable methods (cross-validation or OOB sample reuse) to choose them.

An important and related idea is that of reproducibility.  We encourage researchers to share sufficient code that the entire study could be independently replicated.  In some circumstances (i.e., specialized or commercial software is being used to estimate the model(s) under study), it may only be possible to share the results and the code used to analyze them.  Connecting back to the idea of calibration and fairness, open code makes it possible for others to assess the extent to which methods are fairly compared.

It may not always be easy or even possible to have fair comparisons.  For example, if one was studying power of hypothesis tests as a followup to the KMM study, a real problem is that fair comparison isn't possible unless the type I error is controlled across methods.  We saw in that study that type I error was not entirely controllable in all situations.  Moreover, it would be unrealistic to artificially control type I error rate in a simulation if such control was impossible in real applications.

\section*{References} 
\begin{description}
\item Bingham, D. R., \& Chipman, H. A. (2007). Incorporating Prior Information in Optimal Design for Model Selection, {\it Technometrics}, 49(2), 155–163.
\item Box, G. E., Hunter, W. H., \& Hunter, S. (2005). {\it Statistics for experimenters}, John Wiley \& Sons, New York.
\item Chipman, H. (1996), Bayesian variable selection with related predictors, {\it Canadian Journal of Statistics}, 24: 17-36.
\item Daniel, C. (1959). Use of half-normal plots in interpreting factorial two-level experiments. {\it Technometrics}, 1, 311--341.
\item Gelman, A., Pasarica, C., \& Dodhia, R. (2002). Let's practice what we preach: turning tables into graphs. {\it The American Statistician}, 56, 121--130.
\item Krishnamoorthy, K.,  Mallick, A. \&  Mathew, T. (2011) Inference for the Lognormal Mean and Quantiles Based on Samples With Left and Right Type I Censoring, {\it Technometrics}, 53, 72--83.
\item MacKay, R. J. \& Oldford R. W. (2000) Scientific Method, Statistical Method and the Speed of Light, {\it Statistical Science}, 15, 254--278.
\item Mukerjee, R. \& Wu, C.F. J. {\it A Modern Theory of Factorial Design}, Springer, New York.
\item Nair, V. N., Abraham, B., MacKay, J., Nelder, J. A., Box, G., Phadke, M. S., Kacker, R. N., Sacks, J., Welch, W. J., Lorenzen, T. J., Shoemaker, A. C., Tsui, K. L., Lucas, J. M., Taguchi, S., Myers, R. H., Vining, G. G., \& C. F. J. Wu. (1992). Taguchi’s Parameter Design: A Panel Discussion. {\it Technometrics}, 34(2), 127-–161. 
\item Wu, C.~F.~J. \& Hamada, M.~.S. (2009) {\it Experiments: Planning, Analysis, and Optimization}, 2nd ed., John Wiley \& Sons, New York.
\end{description}

\section{Supplementary Materials}

\subsection{Code}
R code used in the 2 examples is available at \url{https://github.com/hughchipman/TablesAsDesigns}.

\subsection{4-way effects in the KMM analysis} \label{sec:Krish4fi}

\begin{figure}[bph]
  \centering
  \includegraphics[width=\textwidth]{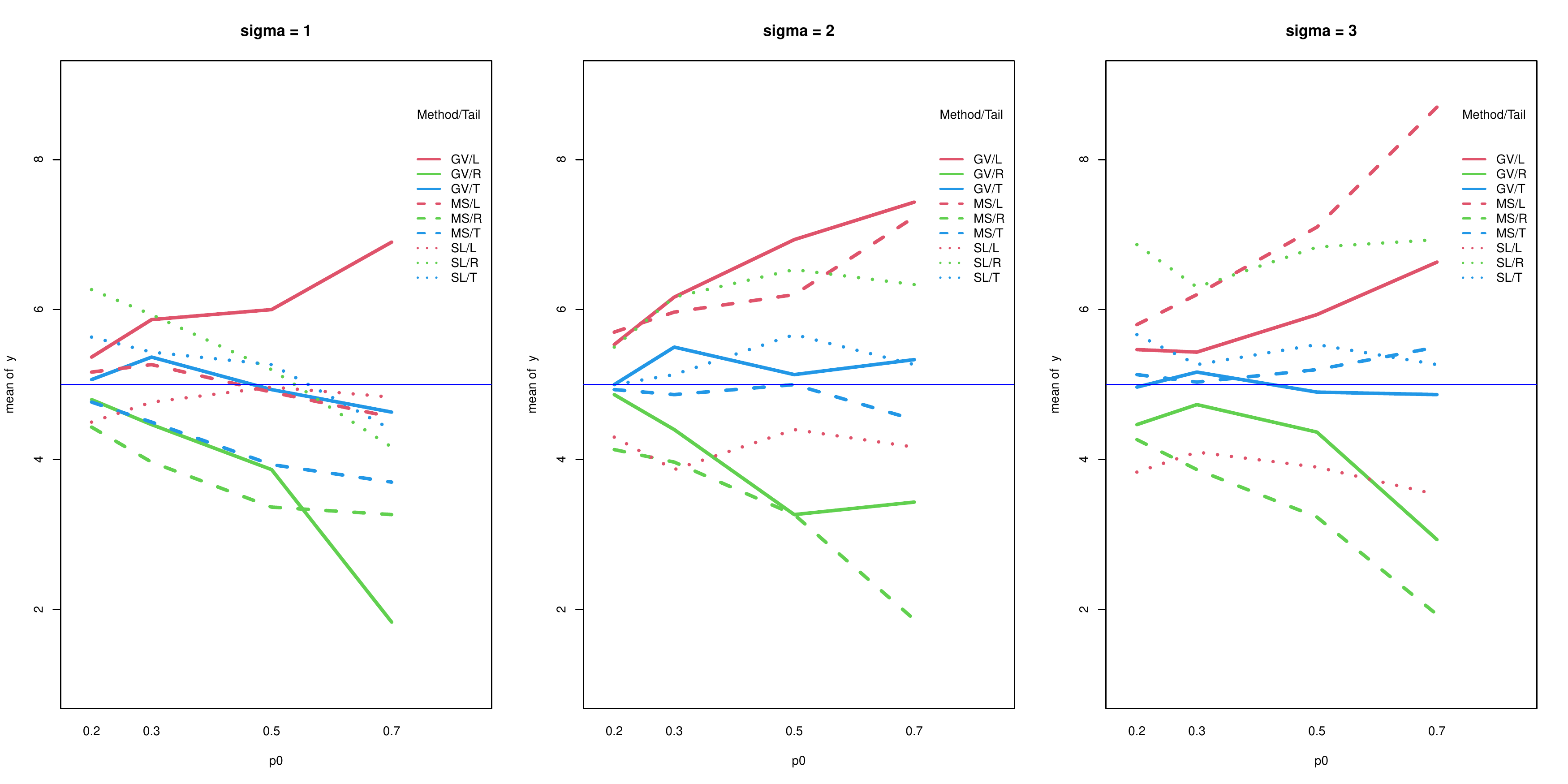}
  \caption{Four-factor interaction plots for Taguchi, showing the effect of {\tt p0} (on the x axes) and {\tt sigma} (across 3 plots). Based on data with {\tt method=AN} excluded. The factor {\tt n} is averaged out, so each point corresponds to 3 table entries ({\tt n = 20, 30, 50}).}
  \label{fig:Krish_Taguchi4fi}
\end{figure}

Interaction plots displaying the joint effects of the 4 factors {\tt method, tail, sigma, p0} are shown in Figure~\ref{fig:Krish_Taguchi4fi}. This is accomplished by dividing up the {\tt method:tail:p0} interaction plot (Figure~\ref{fig:Krish_Taguchi3fi}(a)) up into 3 plots for {\tt sigma = 1, 2, 3}. An alternate visualization would be to switch the roles of {\tt p0} and {\tt sigma}. 

The factor {\tt n} is not displayed in the plot. Each point corresponds to an average across the levels {\tt n = 20, 30, 50}. Sample size {\tt n} is the least important factor in the study, as a model without any effects containing {\tt n} still explains about 89\% of the SS variation in the response.

Here are some interpretations from the plot:
\begin{itemize}
  \item For {\tt method = SL}, the dotted lines are somewhat closer to the desired nominal response level $\alpha = 5\%$. However, the patterns vary with {\tt sigma}. For {\tt sigma=1}, the {\tt SL} tests have type I error that is decreasing with {\tt p0}, or close to level (for the red-left tail). For {\tt sigma = 2, 3}, the level of the {\tt SL} test varies with {\tt tail} but is insensitive to {\tt p0} (dotted lines are nearly horizontal, but change level with {\tt tail}).
  \item The method MS is non-robust for left-tailed tests (red long-dash line in all 3 plots). When {\tt sigma = 1} (left panel), the type I error rate is stable across {\tt p0} and close to 5\%. But for {\tt sigma = 2} or {\tt 3}, the type I error rate is high and furthermore varies across {\tt p0}. 
  \item For {\tt sigma = 3, p0 = 0.7} the {\tt MS} test exhibits the most liberal values seen in the entire study (highest point in {\tt sigma=3} plot corresponds to {\tt MS/L} test).
\end{itemize}

\subsection{Plots for the ``cheapo'' analysis of 72 runs \label{sec:CheapoSup}}

\begin{figure}
  \centering
  \includegraphics[width=\textwidth]{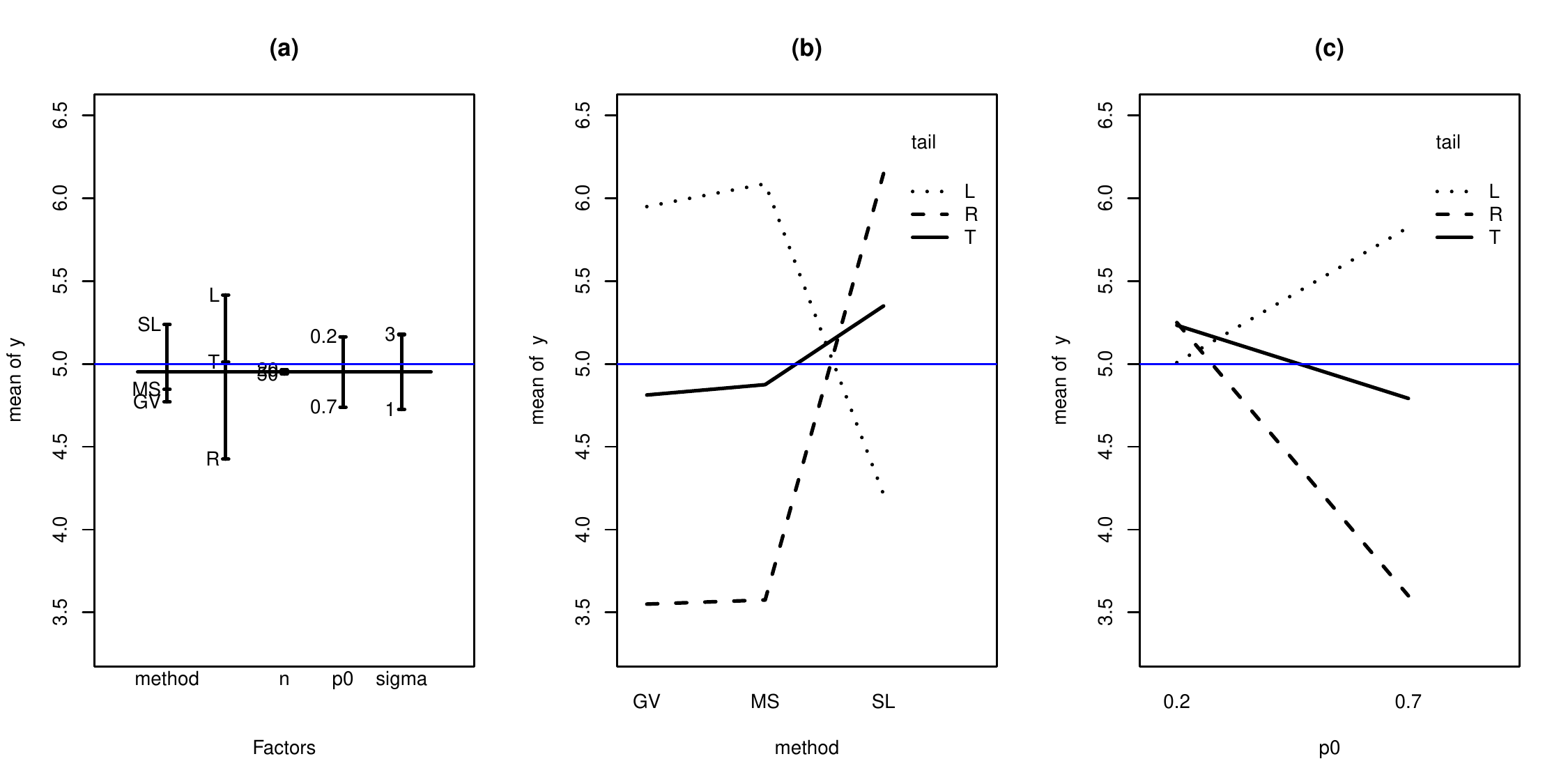}
  \caption{Main effects plot and select two-factor interactions plot. Compare with Figure~\ref{fig:Krish_me2fi}.}
  \label{fig:Krish_cheapo_1}
\end{figure}

\begin{figure}
  \centering
  \includegraphics[width=\textwidth]{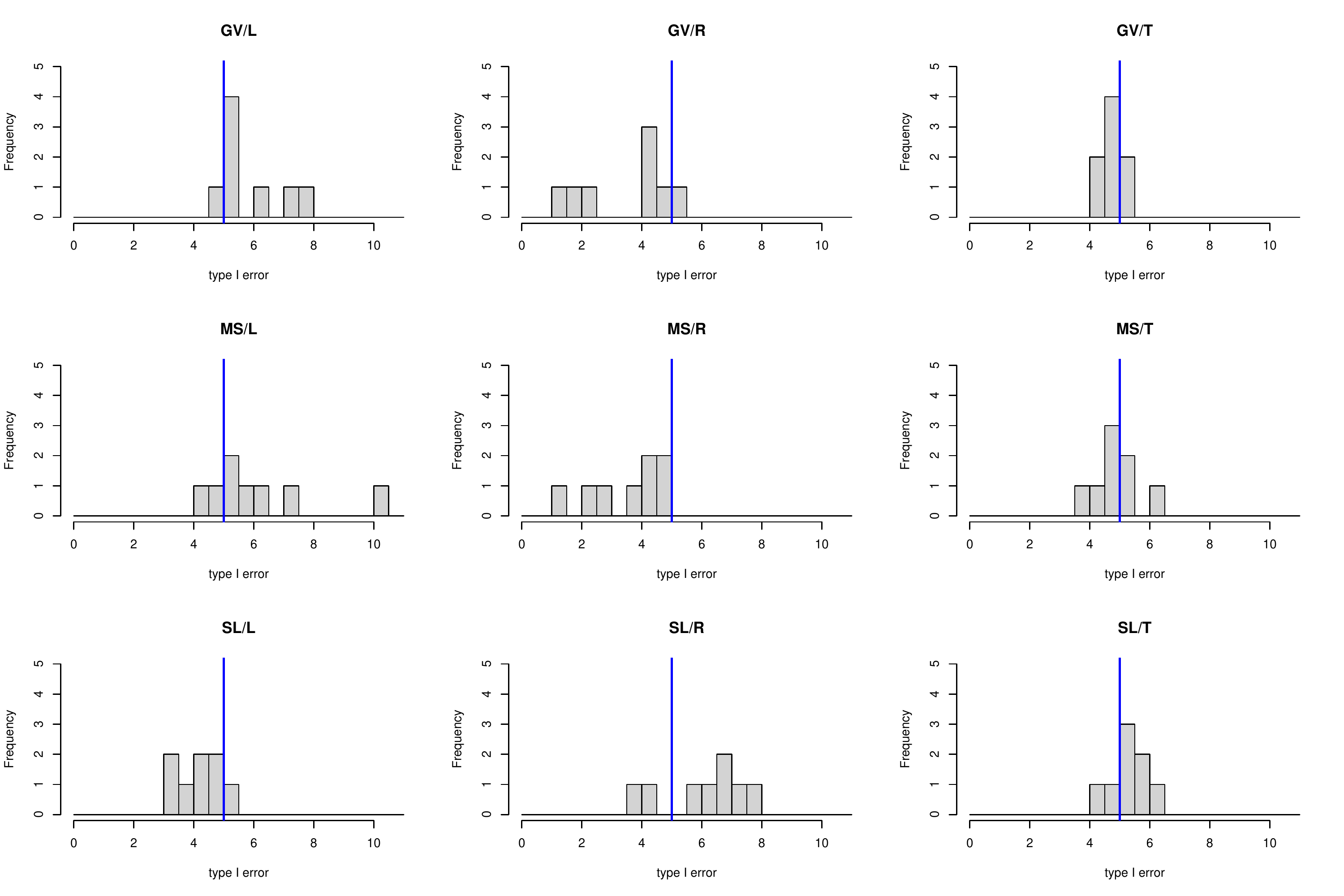}
  \caption{Histograms of response at 9 combinations of the 2 control factors. Compare with Figure~\ref{fig:Krish_TaguchiHist}.}
  \label{fig:Krish_cheapo_2}
\end{figure}

\begin{figure}
  \centering
  \includegraphics[width=\textwidth]{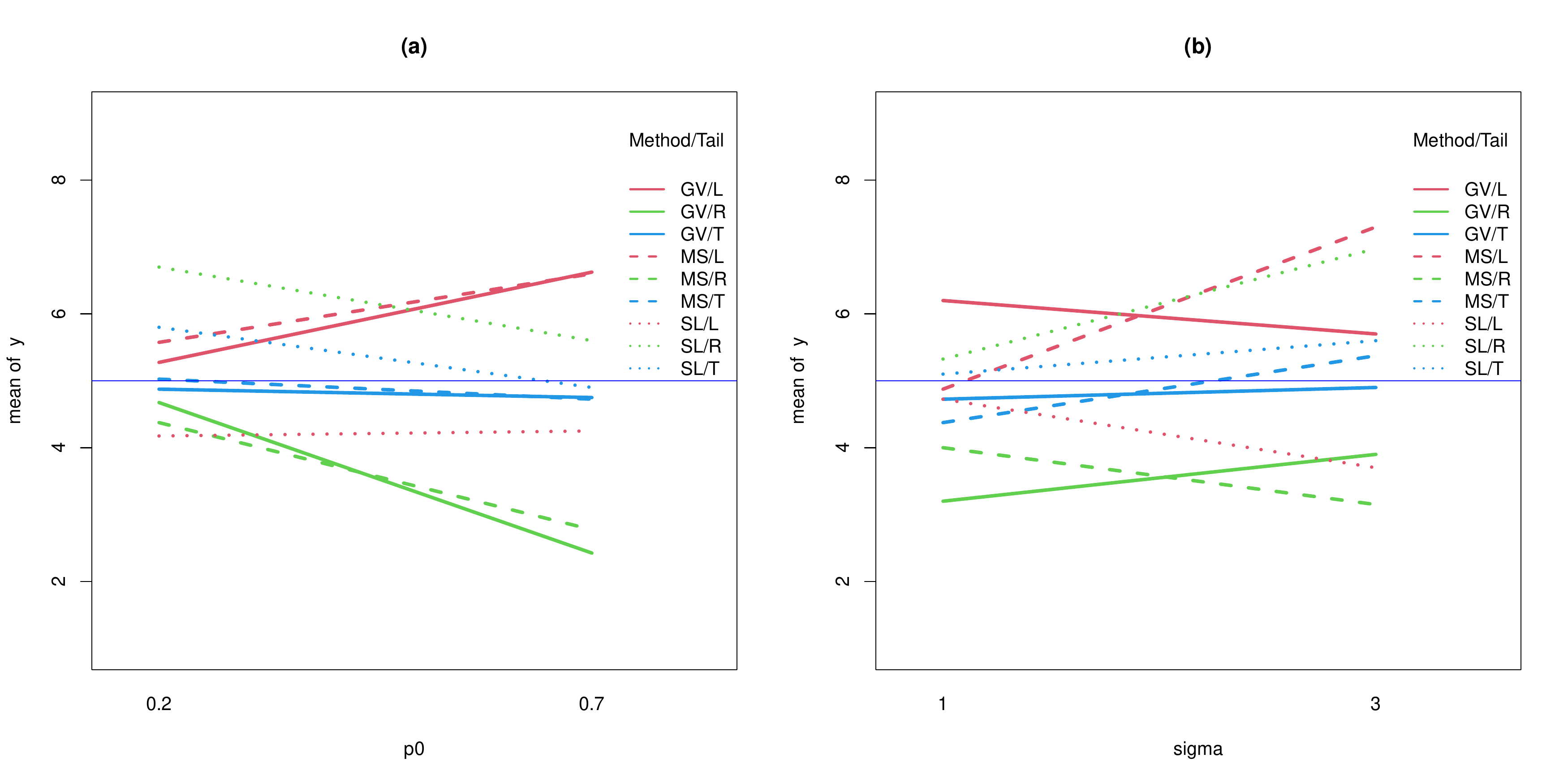}
  \caption{3-way interaction plot. Compare with Figure~\ref{fig:Krish_Taguchi3fi}.}
  \label{fig:Krish_cheapo_3}
\end{figure}

\begin{figure}
  \centering
  \includegraphics[width=\textwidth]{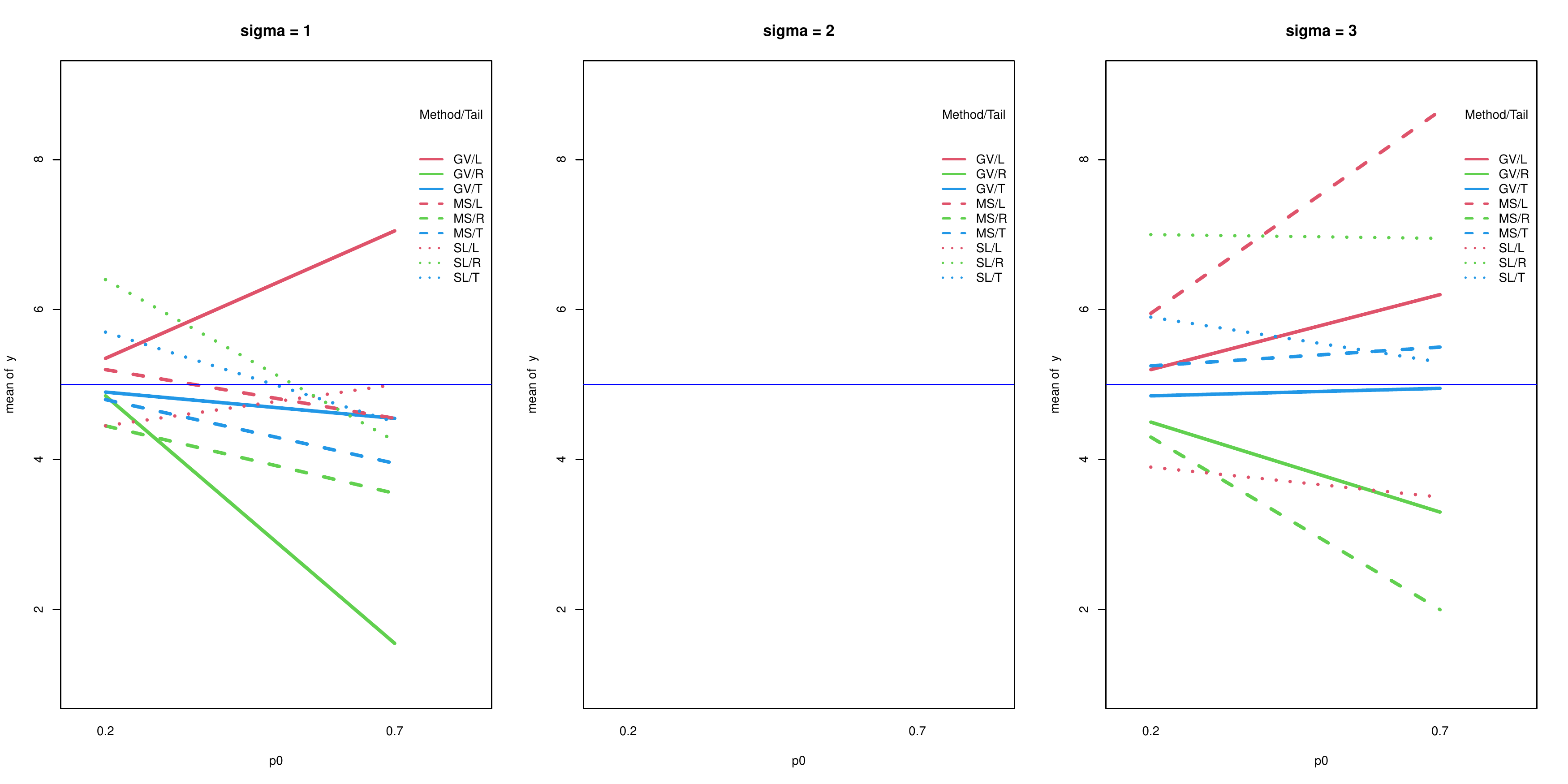}
  \caption{4-way interaction plot. Compare with Figure~\ref{fig:Krish_Taguchi4fi}. }
  \label{fig:Krish_cheapo_4}
\end{figure}

Removing all runs with factors {\tt p0, sigma, n} at values that are not extreme, and deleting {\tt method = AN} runs reduces the original 432 runs to 72 runs. Visualizations of main effects, interactions, and histograms of the response are given in Figures~\ref{fig:Krish_cheapo_1}-\ref{fig:Krish_cheapo_4}. The lead to the same conclusions as a full analysis.


\clearpage

\subsection{Derivations of {\tt ENE} probability distribution \label{sec:ENE_derivations}} 

Main effects are active independently of each other, each with probability $\pi$ of being active. Conditional on the activity of main effects $A$ and $B$, the AB interaction is active with probability
\[
P(AB\mbox{ active } | \mbox{ activity of } A,B)=\left\{\begin{array}{lll}
        0 & \mbox{if both $A, B$ inactive}\\
        c_1\pi & \mbox{if exactly one of $A, B$ active}\\
        c_2\pi & \mbox{if both $A, B$ active}.
                \end{array}\right. .
\]

We follow a similar approach to Appendix B of Bingham and Chipman (2007), calculating a conditional expectation of the number of active effects, given $f$ active main effects and then taking an expectation over $f$. We group terms differently, to identify expected number of effects of certain types.

Conditional on $f$ active main effects, the expected number of active effects is 
\begin{equation} \label{eq:condexp}
  f + f(q-f)c_1\pi + \binom{f}{2}c_2\pi .
\end{equation}
In (\ref{eq:condexp}) the first term corresponds to main effects, the second to 2-factor interactions with 1 parent active, and the last to 2-factor interactions with both parents active. If neither parent is active, then the interaction has 0 probability of being active, and does not contribute to (\ref{eq:condexp}).

With all main effects being active independently and each with probability $\pi$, the number of active main effects $f$ is Binomial with $q$ trials and probability of success $\pi$. Thus $\mbox{E}(f) = \pi q$ and $\mbox{E}(f^2) = \pi q(1-q-\pi q)$. Taking expectation of (\ref{eq:condexp}) with respect to $f$ thus gives, with simplification
\begin{equation} \label{eq:ENEfinal}
  {\tt ENE} = \pi q + c_1\pi^2(1-\pi)q(q-1) + c_2\pi^3q(q-1)/2.
\end{equation}
In (\ref{eq:ENEfinal}), the first, second and third terms correspond to main effects, 2-way interactions with 1 active parent and 2-way interactions with 2 active parents, respectively.

For given values of {\tt ENE} and $q$, and the 50/25/25 split of expected active terms among the 3 parts, the 3 constants $\pi, c_1, c_2$ are the solution of 
$$
  {\tt ENE}/2 = \pi q,
$$
$$
  {\tt ENE}/4 = c_1\pi^2(1-\pi)q(q-1),
$$
and
$$
  {\tt ENE}/4 = c_2\pi^3q(q-1)/2.
$$

Solving the system of equations gives $$\pi = \frac{\tt ENE}{2q}, $$ 
$$c_1 = \frac{\tt ENE}{4\pi^2(1-\pi)q(q-1)}$$
$$c_2 = \frac{\tt ENE}{2\pi^3q(q-1)}$$

\subsection{Analysis of Alternate Designs for the Statistical Learning Experiment}

We consider 4 scenarios: 
\begin{enumerate}
  \item Original full-factorial design with 2 replicates (256 runs)
  \item Full-factorial design without replication (128 runs)
  \item Quarter fraction design with 2 replicates (64 runs)
  \item Quarter fraction design without replication (32 runs, as analyzed in Section~5.1
\end{enumerate}

\begin{figure}
  \centering
  \includegraphics[width=\textwidth]{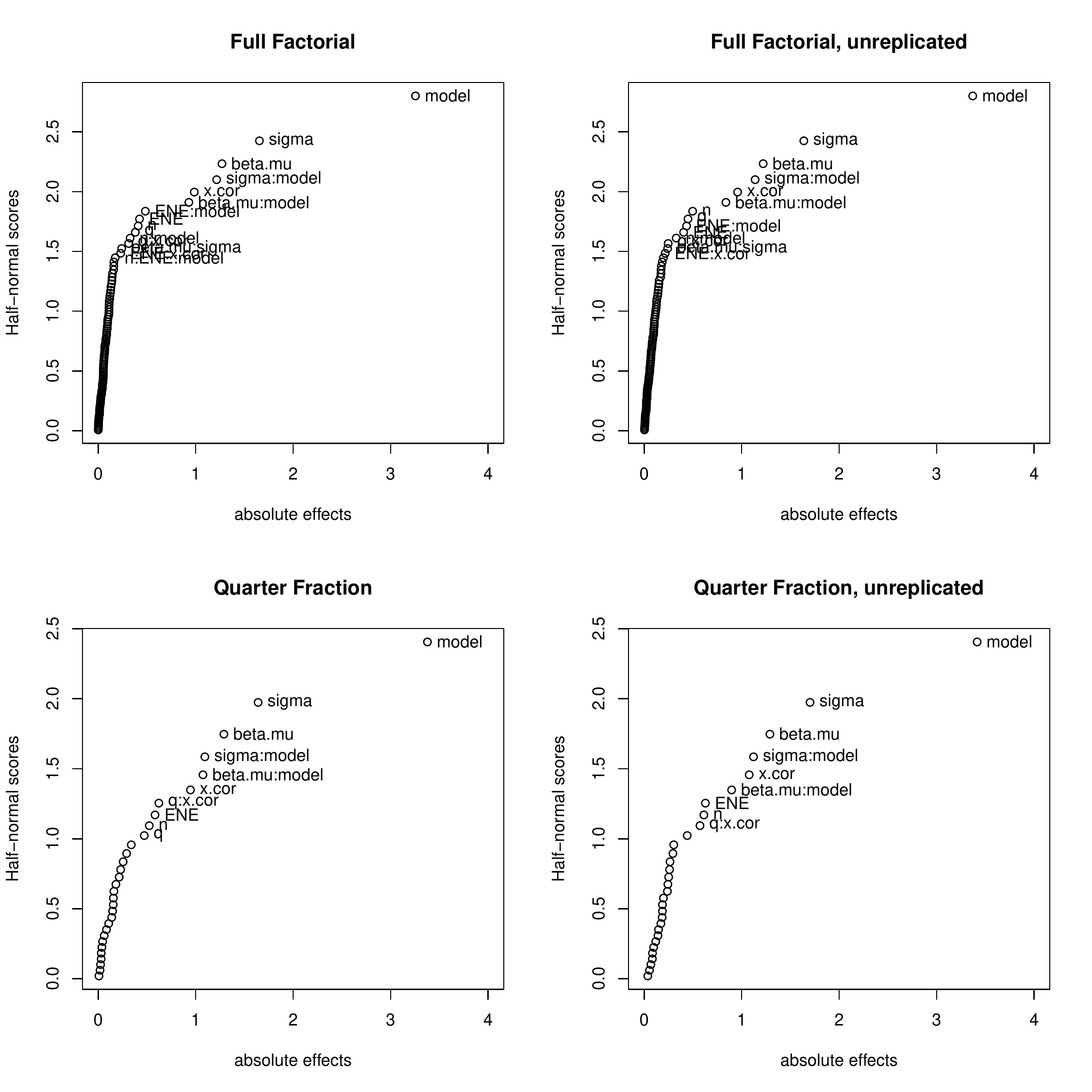}
  \caption{Half-normal plots of effects for the 4 different designs. Large effects are those on the right. The plot for the quarter fraction unreplicated experiment is the same as in Section 5.1. 
  }
  \label{fig:super_halfnormal}
\end{figure}

Figure~\ref{fig:super_halfnormal} displays the half-normal plots for the 4 scenarios. The 6 largest effects {\tt model}, {\tt sigma}, {\tt beta.mu}, {\tt sigma:model}, {\tt x.cor} and {\tt beta.mu:model} are the same for all plots, and these constitute the main elements of the original analysis given in Section~5.1. 
Some smaller effects that still could be significant are aliased, such as {\tt q:x.cor}. This effect is aliased with {\tt ENE:beta.mu} in the two quarter-fraction designs (bottom row of plots), and these designs do not allow the effects to be disentangled.

Since the same effects are large, and the estimates are similar across the 4 different experiments, we do not duplicate earlier analysis. A full analysis would include interpretation of main effects plots and interaction plots, and consideration of TRPD. Conclusions in this case will be the same as before, and because the largest effects are not aliased with other low-order terms, we do not lose any information.

Analysis based on an ANOVA table would also be possible. For the 2 replicated designs (64 and 256 runs), there are degrees of freedom for error, even if all possible effects are estimated. In unreplicated designs, some higher-order interactions would need to be assumed to be zero to allow degrees of freedom for an error sum of squares. 
Having already demonstrated for this example that the analyses are very similar for all 4 designs, ANOVA tables are not presented here.

Replication of the design (e.g. running 2 replicates at each of the 32 settings of the quarter fraction) would give a ``pure error'' estimate of residual standard deviation in the ANOVA model. Running a full factorial or a design with less fractionation (e.g., here, a half-fraction) would eliminate or reduce aliasing of effect estimates.

\clearpage

\end{document}